\documentclass{article}
\usepackage{amsmath,amssymb,amsfonts,graphicx,bm}
\usepackage{pgfplots}

\newcommand{\e}{\mathrm e}
\newcommand{\n}{\mathbf n}
\newcommand{\x}{\mathbf{x}}
\newcommand{\z}{\mathbf{z}}

\newcommand{\y}{\mathbf{y}}

\newcommand{\X}{\mathbf{X}}
\renewcommand{\P}{\mathbb{P}}
\newcommand{\p}{\widetilde{p}}

\newcommand{\dis}{\displaystyle}

\def\E{\mathbb{E}}
\def\R{\mathbb{R}}

\def\P{\mathbb{P}}
\def\L{\mathbb{L}}

\def\calU{\mathcal{U}}

\def\calA{\mathcal{A}}

\def\calV{\mathcal{V}}
\def\calT{\mathcal{T}}

\def\PP{\widetilde{\mathcal{P}}}
\def\Q{\widetilde{Q}}
\def\X{\mathbf{X}}
\def\calF{\mathcal{F}}

\begin{document}

\title{Semipermeable interfaces and the target problem}
\author{\em Paul C. Bressloff,\\
Department of Mathematics, Imperial College London, \\
London SW7 2AZ, UK.}
\maketitle

\begin{abstract}
In this chapter, we review our recent work on first passage time  (FPT) problems for absorption by a target whose interface is semipermeable. For pedagogical reasons, we focus on a single Brownian particle searching for a single target in a bounded domain. We begin by writing down the forward diffusion equation for the target problem, and define various quantities of interest such as the survival probability, absorption flux, and the FPT density. We also present a general method of solution based on Green's functions and the spectral decomposition of so-called Dirichlet-to-Neumann (D-to-N) operators. We then use an encounter-based approach to extend the theory to the case of non-Markovian absorption within the target interior. Encounter-based models consider the joint probability density or generalized propagator for particle position and the amount of particle-target contact time prior to absorption. In the case of a partially absorbing target interior, the contact time is given by a Brownian functional known as the occupation time. Finally, we develop a more general probabilistic model of single-particle diffusion through semi-permeable interfaces, by combining the encounter-based approach with so-called snapping out Brownian motion (BM).  Snapping out BM sews together successive rounds of partially absorbing BMs that are restricted to either the interior or the exterior of the semipermeable interface. The rule for terminating each round is implemented using encounter-based model of partially absorbing BM. We show that this results in a time-dependent permeability that can be heavy-tailed.

\end{abstract}

\section{Introduction}

\begin{figure}[t!]
\centering
  \includegraphics[width=10cm]{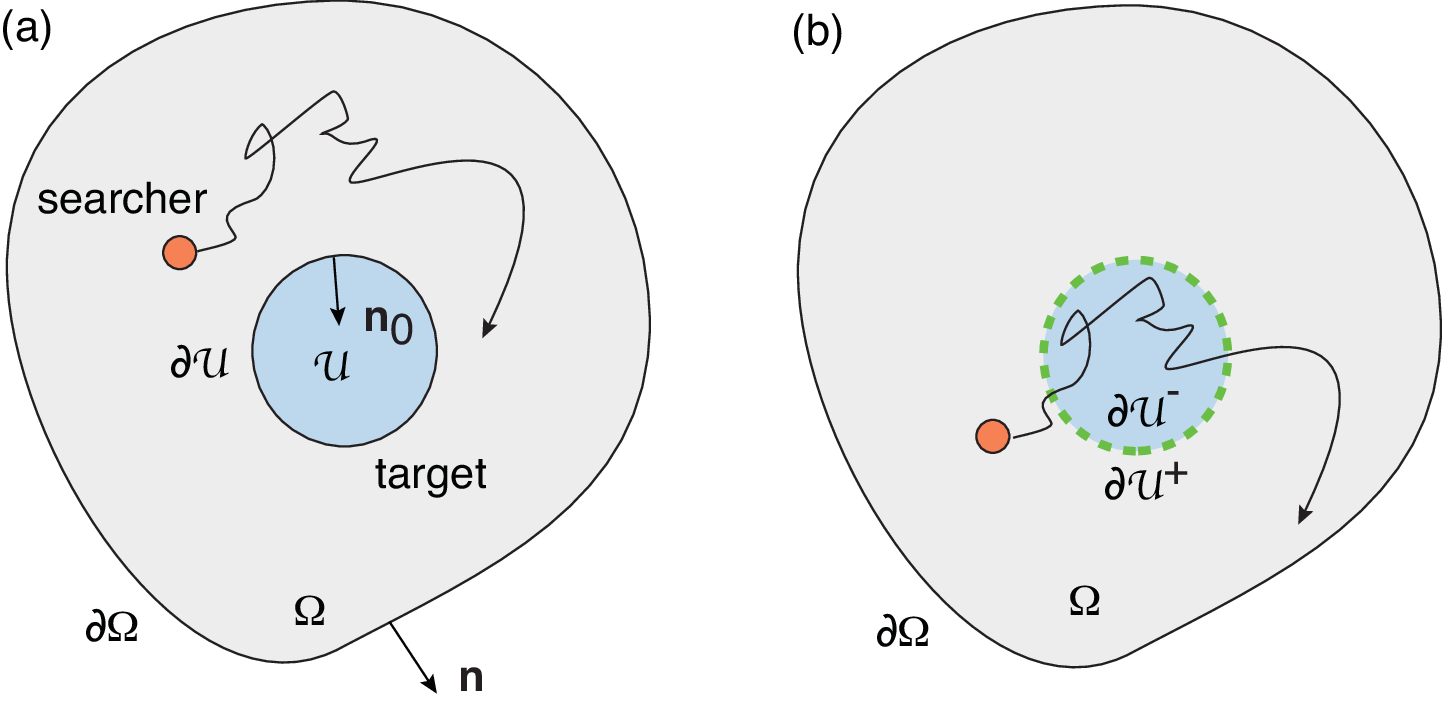}
  \caption{Diffusion of a particle (searcher) in a bounded domain $\Omega$ with $\calU\subset \Omega\subset \R^d$. (a) Partially reactive interface $\partial \calU$ in which the probability of absorption depends on the particle-surface encounter time (boundary local time). (b) Semi-permeable interface $\partial \calU$ with a partially absorbing target interior or trap $\calU$. The particle can now enter and exit the target. The probability flux across $\partial \calU$ is continuous, but there is a jump discontinuity in the probability density that depends on the permeability of $\partial \calU$. The probability of particle absorption depends on the amount of time spent within $\calU$ (occupation time).}
  \label{fig1}
\end{figure}

A classical example of a target problem is shown in Fig. \ref{fig1}. A Brownian particle or searcher is confined within some bounded domain $\Omega \subset \R^d$ that contains an interior target $\calU \subset \Omega$. Assuming that the exterior boundary $\partial \Omega$ is totally reflecting, one is typically interested in calculating the statistics of the first passage time (FPT) for the particle to be absorbed by (find) the target. The details will depend on the nature of the target and its surface interface $\partial \calU$. The most common scenario is shown in Fig. \ref{fig1}(a), where $\partial \calU$ is a partially reactive surface. That is, whenever the particle encounters the target boundary, it is absorbed at some constant rate $\kappa_0$ or reflected back into the domain $\Omega\backslash \calU$. (In the limit $\kappa_0\rightarrow \infty$, the interface $\partial \calU$ becomes totally absorbing.) An alternative scenario is illustrated in Fig. \ref{fig1}(b), where $\partial \calU$ now acts as a semipermeable membrane surrounding a partially absorbing target $\calU$. The particle flux across the interface is continuous but there is a jump discontinuity in the density. Whenever the particle diffuses within $\calU$, it is absorbed at some constant rate $\gamma$. One notable example of the latter scenario is the lateral diffusion of neurotransmitter receptors within the plasma membrane of a neuron. The partially absorbing traps correspond to local synaptic trapping regions that bind receptors to scaffolding proteins, followed by internalization of the receptors via endocytosis \cite{Earnshaw06,Holcman06,Bressloff08,Thoumine12,Schumm22,Bressloff23a}. Treating the synaptic interfaces as semi-permeable membranes is motivated by the so-called partitioned fluid-mosaic model of the plasma membrane \cite{Kusumi05}, in which confinement domains are formed by a fluctuating network of cytoskeletal fence proteins combined with transmembrane picket proteins that act as fence posts. 

In this chapter, we review our recent work on FPT problems for absorption by a target $\calU$ whose interface $\partial \calU$ is semipermeable. For pedagogical reasons, we focus on a single Brownian particle searching for a single target in a bounded domain. We begin by writing down the forward diffusion equation for the target problem shown in Fig. \ref{fig1}(b), and define various quantities of interest such as the survival probability, absorption flux, and the FPT density (see section 2). We then present a general method of solution based on Green's functions and the spectral decomposition of so-called Dirichlet-to-Neumann (D-to-N) operators. Such methods have previously been applied to the target problem of Fig. \ref{fig1}(a) \cite{Grebenkov19b,Grebenkov20} and the target problem of Fig. \ref{fig1}(b) when the interface $\partial \calU$ is transparent \cite{Bressloff22a}. In the multidimensional case ($d>1$), $\partial \calU$ is a finite-dimensional compact surface and the D-to-N operators have countably infinite spectra. This means that the solution of the diffusion equation takes the form of an infinite series that requires inverting an infinite-dimensional matrix. In section 4, we consider the simpler problem of diffusion in the interval $\Omega=[-L',L]$ with a partially absorbing subinterval $\calU=[-L',0)$ and a semipermeable membrane at $x=0$. The D-to-N operators reduce to scalar multipliers, which allows us to derive an explicit formula for the mean FPT (MFPT) as a function of the interface permeability.

In section 5, we use an encounter-based approach to extend the theory of partial absorption within the target $\calU$. Encounter-based models consider the joint probability density or generalized propagator for particle position and the amount of particle-target contact time prior to absorption \cite{Grebenkov20,Bressloff22a,Grebenkov22,Bressloff22}. Absorption occurs when the contact time exceeds a random threshold. If the probability distribution of the latter is an exponential function, then one recovers the Markovian example of absorption at a constant rate, whereas a non-exponential distribution signifies non-Markovian absorption. In the case of a partially absorbing target $\calU$ (surface $\partial \calU$) the contact time is given by a Brownian functional known as the occupation time (boundary local time). We illustrate the theory using the 1D example of section 4. In particular, we derive an explicit expression for the MFPT that depends on various moments of the occupation time threshold distribution. Finally, in section 6, we  develop a more general probabilistic model of single-particle diffusion through semi-permeable interfaces, by combining the encounter-based approach with so-called snapping out Brownian motion (BM) \cite{Bressloff22b,Bressloff23,Bressloff23b}. The latter was originally formulated for 1D single-particle diffusion through a semipermeable barrier \cite{Lejay16,Lejay18,Brobowski21}, but has recently been extended to higher spatial dimensions \cite{Bressloff23}. Snapping out BM sews together successive rounds of partially absorbing BMs that are restricted to either the interior or the exterior of $\partial \calU$. The rule for terminating each round is implemented using the encounter-based model
of partially absorbing BM introduced in Ref. \cite{Grebenkov20}. We show that this results in a time-dependent permeability that can be heavy-tailed.

\section{Single target with a semipermeable interface}

Consider the single target problem shown in Fig. \ref{fig1}(b), in which a semipermeable interface $\partial \calU$ surrounds a partially absorbing interior $\calU$, with $\partial \calU^+ $ ($\partial \calU^-$) denoting the side approached from outside (inside) $\calU$.
Let $p(\x,t|\x_0)$ denote the probability density that the particle position $\X_t$  is in a neighborhood of $\x\in \Omega\backslash\calU$ at time $t$, given that it started at $\x_0$. That is,
$p(\x,t|\x_0)d\x=\P[\x<\X_t<\x+d\x|\X_0=\x_0]$.
Denote the corresponding probability density within $\calU$ by $q(\x,t|\x_0)$. The forward diffusion equation takes the form
\begin{subequations} 
\label{master}
\begin{eqnarray}
	\frac{\partial p(\x,t|\x_0)}{\partial t} &=& D\nabla^2 p(\x,t|\x_0), \quad \x\in \Omega \backslash \calU,\quad \nabla p\cdot \n=0,\quad \x\in \partial \Omega,\\
	\frac{\partial q(\x,t|\x_0)}{\partial t} &=& D\nabla^2 q(\x,t|\x_0) -\gamma q(\x,t|\x_0),\quad \x\in \calU,
	\end{eqnarray}
where $\gamma$ is the rate at which the particle is absorbed within the target $\calU$. These are supplemented by the 
semipermeable boundary conditions
\begin{align}
 D\nabla p(\x^+,t|\x_0)\cdot \n_0&= D\nabla q(\x^-,t|\x_0)\cdot \n_0\equiv -  {\mathcal J}(\x,t|\x_0) \\
	{\mathcal J}(\x,t|\x_0)&=\kappa\bigg [(1-\alpha )p(\x^+,t|\x_0)- \alpha q(\x^-,t|\x_0)\bigg ] ,\quad \x^{\pm}  \in \partial \calU^{\pm},
\end{align}
\end{subequations}
\noindent where ${\mathcal J}(\x,t|\x_0)$ is the continuous inward flux across the point $\x\in \partial \calU$, $\kappa$ is the permeability of the interface $\partial \calU$, and $\alpha\in [0,1]$ specifies a directional bias with $\alpha=1/2$ the unbiased case. 
(One could also take the diffusivities within $\Omega\backslash \calU$ and $\calU$ to be different.) Equations (\ref{master}c) and (\ref{master}d) are one version of the well-known Kedem-Katchalsky (KK) equations \cite{Kedem58,Kedem62,Kargol96}. Note that in the limit $\kappa \rightarrow \infty$ with $\alpha=1/2$, the interface is transparent and we obtain the pair of  continuity equations
\begin{align}
\label{cont}
 D\nabla p(\x^+,t|\x_0)\cdot \n_0&= D\nabla q(\x^-,t|\x_0)\cdot \n_0,\quad  p(\x^+,t|\x_0)=q(\x^-,t|\x_0)  \end{align}
for $\x^{\pm}  \in \partial \calU^{\pm}$
On the other hand, if $\kappa=0$, then the interface is totally reflecting on both sides. Finally, if the particle started outside the domain $\calU$, then in the limit $\gamma \rightarrow \infty$ the particle is absorbed as soon as it hits the target boundary, see Fig. \ref{fig1}(a). Hence, we recover a totally absorbing target with $p$ evolving according to equations (\ref{master}a) such that $p(\x,t|\x_0)=0$ for all $\x\in \partial \calU$.

Consider the survival probability that the particle hasn't been absorbed by the target in the time interval $[0,t]$, having started at $\x_0$ \cite{Redner01}:
\begin{equation}
\label{Qin}
 S(\x_0,t)=\int_{\Omega\backslash \calU}p(\x,t|\x_0)d\x+ \int_{\calU}q(\x,t|\x_0)d\x.
\end{equation}
Differentiating both sides of this equation with respect to $t$ and using equations (\ref{master}a)-(\ref{master}d) together with the divergence theorem gives
\begin{eqnarray}
\frac{\partial S(\x_0,t)}{\partial t}&=&D \int_{\partial \calU^+} \nabla p(\x,t|\x_0)\cdot \n_0 d\sigma-D\int_{\partial \calU^-} \nabla q(\x,t|\x_0)\cdot \n_0 d\sigma\nonumber \\
&&\quad  -\gamma  \int_{\calU} q(\x,t|\x_0)d\x.
\end{eqnarray}
Continuity of the flux across the interface implies that the first two terms on the right-hand side cancel, so that
\begin{equation}
\frac{\partial S(\x_0,t)}{\partial t}= -\gamma  \int_{\calU} q(\x,t|\x_0)d\x \equiv -J(\x_0,t),
\label{Qin2}
\end{equation}
where $J(\x_0,t)$ is the total absorption flux within $\calU$. It follows that we can identify $J(\x_0,t)$ with the FPT density $f(\x_0,t)$. In particular, the moments of the FPT density can be written as
\begin{eqnarray}
\E[\calT^n]&=&\int_0^{\infty}J(\x_0,t)t^ndt =\lim_{s\rightarrow 0}\left (-\frac{\partial }{\partial s}\right )^n\int_0^{\infty} \e^{-st}J(\x_0,t)dt\nonumber \\
&=&\lim_{s\rightarrow 0}\left (-\frac{\partial }{\partial s}\right )^n\widetilde{J}(\x_0,s),
\end{eqnarray}
where $\widetilde{J}(\x_0,s)$ is the Laplace transformed flux. In other words, the latter acts as the moment generating function for the FPT density. 
Finally, the Laplace transformed fluxes $\widetilde{J}(\x_0,s)$ and $\widetilde{\mathcal J}(\x,t|\x_0)$ can be related as follows. First, equation (\ref{Qin2}) implies
\begin{align}
\frac{\partial J(\x_0,t)}{\partial t}&= \gamma  D\int_{\calU}\nabla^2 q(\x,t|\x_0) d\x-\gamma J(\x_0,t) \nonumber \\
&=\gamma\int_{\partial \calU} {\mathcal J}(\x,t|\x_0) d\x-\gamma J(\x_0,t),
\label{Jint}
\end{align}
where ${\mathcal J}(\x,t|\x_0)= - \nabla q(\x,t|\x_0) \cdot \n_0 $.
Laplace transforming equation (\ref{Jint}) with respect to $t$ and using the initial condition $J(\x_0,0)=0$ for $\x_0\notin \calU$, we have
\begin{equation}
(s+\gamma) \widetilde{J}(\x_0,s)=\gamma \int_{\partial \calU} \widetilde{ {\mathcal J}}(\x,s|\x_0) d\x.
\end{equation}
In particular, for all $\x_0 \notin \calU$
\begin{align}
\label{TRobin}
T(\x_0)\equiv -\left . \frac{\partial}{\partial s}\widetilde{J}(\x_0,s)\right |_{s=0}= \frac{1}{\gamma} - \int_{\partial \calU} \partial_s\widetilde{ {\mathcal J}}(\x,0|\x_0) d\x .
\end{align}
Finally, in order to determine $\widetilde{J}(\x_0,s)$, it is necessary to solve the forward diffusion equation in Laplace space:
\begin{subequations}
\label{LTmaster}
\begin{align}
& D\nabla^2 \widetilde{p}(\x,s|\x_0)-s \widetilde{p}(\x,s|\x_0)=-\delta(\x-\x_0),\ \x,\x_0 \in \Omega\backslash \calU , \\
&-\nabla  \widetilde{p}(\x,s|\x_0) \cdot \n=0,\  \x\in \partial \Omega,\\
 &D\nabla^2  \widetilde{q}(\x,s|\x_0)-(s+\gamma) \widetilde{q}(\x,s|\x_0) =0,\, \x\in \calU \\
&
 D\nabla \widetilde{p}(\x^+,s|\x_0)\cdot \n_0= D\nabla \widetilde{q}(\x^-,s|\x_0)\cdot \n_0\equiv -  \widetilde{\mathcal J}(\x,s|\x_0) \\
&	\widetilde{\mathcal J}(\x,s|\x_0)=\kappa\bigg [(1-\alpha )\widetilde{p}(\x,s|\x_0)- \alpha \widetilde{q}(\x,s|\x_0)\bigg ] ,\quad \x\in \partial \calU.
 \end{align}
 \end{subequations}
 For the sake of illustration, we have taken $\x_0\in \Omega\backslash \calU$.

\setcounter{equation}{0}
\section{Green's functions and spectral decompositions}

As highlighted in the previous section, one way to calculate the moments of the FPT density for a partially absorbing target $\partial \calU$ is to solve the forward diffusion equation in Laplace space, which yields the Laplace transformed target flux $\widetilde{J}(\x_0,s)$. In the case of a partially absorbing interface $\partial \calU$, a general method for solving the corresponding Robin BVP is based on a spectral decomposition of a so-called Dirichelt-to-Neumann (D-to-N) operator \cite{Grebenkov20}. 
The analogous spectral analysis for a semi-permeable interface $\partial \calU$ is considerably more involved when $d\geq 2$. This is true even in the infinite permeability limit $\kappa \rightarrow \infty$ with $\alpha=1/2$, for which the interface $\calU$ becomes completely transparent. The latter example was analyzed in Ref. \cite{Bressloff22a} by replacing the continuity equations (\ref{cont}) with the inhomogeneous Dirichlet conditions (in Laplace space) $\widetilde{p}(\x,s|\x_0)=\widetilde{q}(\x,s|\x_0)=f(\x,s)$ for all $\x\in \partial \calU$. Here we consider the case of finite $\kappa$. The first step is to replace the semipermeable boundary conditions (\ref{master}d,e) with a pair of Dirichlet conditions
$\widetilde{p}(\x,s|\x_0)=f(\x,t)$ and $\widetilde{q}(\x,s|\x_0)=\overline{f}(\x,s)$ for the unknown functions $f,\overline{f}$. The general solution of equations (\ref{LTmaster}a)--(\ref{LTmaster}c) can then be written in the form
  \begin{align}
 \label{sool}
 \widetilde{p}(\x,s|\x_0)&= \calF(\x,s)+G(\x,s|\x_0),\ \x \in \Omega \backslash \calU,\quad   \widetilde{q}(\x,s|\x_0)= \overline{\calF}(\x,s),\ \x \in \calU,
 \end{align}
 where 
  \begin{subequations}
 \begin{align}
  \calF(\x,s)&=-D\int_{\partial \calU} \partial_{\sigma'} G(\x',s|\x)f(\x',s)d\x',\\ \overline{\calF}(\x,s)&=D\int_{\partial \calU} \partial_{\sigma'} \overline{G}(\x',s+\gamma|\x)\overline{f}(\x',s)d\x',
  \end{align}
  \end{subequations}
  and $\partial_{\sigma'}=\nabla_{\x'}\cdot \n_0$.
We have introduced the modified Helmholtz Green's functions $G$ and $\overline{G}$ for the two domains $\calU^c=\Omega\backslash \calU$ and $\calU$, respectively: 
   \begin{subequations}
   \label{nabG}
  \begin{align}
 &D\nabla^2 G(\x,s|\x')-sG(\x,s|\x')=-\delta(\x-\x'),\ \x,\x' \in \Omega\backslash \calU,\\
 &G(\x,s|\x')=0,\ \x\in \partial \calU, \quad \nabla G(\x,s|\x')\cdot \n=0,\ \x\in \partial \Omega,\\
& D\nabla^2 \overline{G}(\x,s|\x')-s\overline{G}(\x,s|\x')=-\delta(\x-\x'),\ \x,\x' \in \calU , \\
&\overline{G}(\x,s|\x')=0,\ \x \in \partial \calU.
\end{align}
\end{subequations}
The Green's functions have dimensions of [time]/[Length]$^{d}$

The unknown functions $f,\overline{f}$ are determined by substituting the solutions (\ref{sool}a,b) into equations (\ref{LTmaster}d,e):
\begin{subequations}
\label{fL0}
\begin{align}
\L_s[f](\x,s)+\partial_{\sigma}G(\x,s|\x_0)&=-\overline{\L}_{s+\gamma}[\overline{f}](\x,s),\\
\L_s[f](\x,s)+\partial_{\sigma}G(\x,s|\x_0)&=-\frac{\kappa}{D}\bigg [(1-\alpha )f(\x,s)- \alpha \overline{f}(\x,s)\bigg ],\quad \x \in \partial \calU,
\end{align}
\end{subequations}
where $\L_s$ and $\overline{\L}_s$ are the D-to-N operators
\begin{subequations}
\label{DtoN2}
\begin{align}
\L_s[f](\x,s) &=-D\partial_{\sigma}\int_{\partial \calU}\partial_{\sigma'}G(\x',s|\x)f(\x',s)d\x',\\ \overline{\L}_s[\overline{f}](\x,s) &=-D\partial_{\sigma}\int_{\partial \calU}\partial_{\sigma'}\overline{G}(\x',s|\x)\overline{f}(\x',s)d\x'.
\end{align}
\end{subequations}
acting on the space $L_2(\partial \calU)$. The D-to-N operators $\L_s$ and $\overline{\L}_s$ both have discrete spectra. That is, there exist countable sets of eigenvalues $\lambda_n(s),\overline{\lambda}_n(s)$ and eigenfunctions $v_n(\x,s),\overline{v}_n(\x,s)$ satisfying (for fixed $s$)
\begin{equation}
\label{eig}
\L_s v_n(\x,s)=\lambda_n(s)v_n(\x,s),\quad \overline{\L}_s \overline{v}_n(\x,s)=\overline{\lambda}_n(s)\overline{v}_n(\x,s).
\end{equation}
We can now solve equations (\ref{fL0}) by introducing the eigenfunction expansions 
\begin{equation}
\label{eig2}
f(\x,s)=\sum_{m=0}^{\infty}f_m(s) v_m(\x,s),\quad \overline{f}(\x,s)=\sum_{m=0}^{\infty}\overline{f}_m(s) {v}_m(\x,s)
\end{equation}
Substituting equation (\ref{eig2})
into (\ref{fL0}) and taking the inner product with the adjoint eigenfunctions $v_n^*(\x,s)$ yields the following matrix equations for the coefficients $f_m,\overline{f}_m$:
\begin{subequations}
\label{sspec0}
\begin{align}
\lambda_n(s)f_n(s) -g_n(s)&=-\sum_{m\geq 1} H_{nm}(s+\gamma)\overline{f}_m(s),\\
\lambda_n(s)f_n(s) -g_n(s)&=-\kappa\bigg [(1-\alpha ) f_n(s)- \alpha \overline{f}_n(s) \bigg ]
\end{align}
\end{subequations}
where 
\begin{subequations}
\begin{align}
\label{g}
g_n(s)&=-\int_{\partial \calU}v_n^*(\x,s) \partial_{\sigma}G(\x,s|\x_0)d\x=\frac{1}{D}\calV_n^*(\x_0,s),\\
\label{H}
\calF_n(s)&=D\int_{\partial \calU} v_n^*(\x,s) {\mathcal V}_n(\x,s)d\x,\quad 
\overline{\calF}_n(s)=D\int_{\partial \calU} v_n^*(\x,s) \overline{\mathcal V}_n(\x,s)d\x,\\
H_{nm}(s)&=D\int_{\partial \calU} v_n^*(\x,s)\partial_{\sigma}{\mathcal V}_m(\x,s)d\x.
\end{align}
\end{subequations}
Here $\calV_n$ and $\overline{\calV}_n$ are defined according to
\begin{subequations}
\label{Vn}
  \begin{align}
{\mathcal V}_n(\x,s)&=-D\int_{\partial \calU} v_n(\x',s)\partial_{\sigma'}G(\x',s|\x)d\x',\\
\overline{\mathcal V}_n(\x,s)&=D\int_{\partial \calU} v_n(\x',s)\partial_{\sigma'}\overline{G}(\x',s|\x)d\x'.
\end{align}  
 \end{subequations}
 
Equation (\ref{sspec0}b) implies that
\begin{equation}
\overline{ f}_n(s)=\frac{1}{\kappa \alpha }\bigg [(\lambda_n(s)+\kappa(1-\alpha))f_n(s)-g_n(s)\bigg ].
\end{equation}
Introducing the vectors ${\bf f}(s)=(f_n(s), n\geq 0)$ and ${\bf g}(s)=(g_n(s), n\geq 0)$, we can now formally write the solution of equation (\ref{sspec0}a) as
\begin{align}
{\bf f}(s)&=\left [{\bf M}(s)+\frac{1}{\kappa \alpha }{\bf H}(s+\gamma)\left [{\bf M}(s)+\kappa(1-\alpha){\bf I}\right ]\right ]^{-1} \left [{\bf I}+\frac{1}{\kappa \alpha} {\bf H}(s+\gamma)\right ]{\bf g}(s)\nonumber \\
&\equiv {\bm \Lambda}^{-1}_1(s){\bm \Lambda}_2(s){\bf g}(s),
\label{eff}
\end{align}
where ${\bf H}(s)$ is the matrix with elements $H_{nm}(s)$ and ${\bf M}(s)=\mbox{diag}(\lambda_1(s),\lambda_2(s)\ldots)$. Finally, substituting equation (\ref{eff}) into equations (\ref{sool}) gives
\begin{subequations}
 \label{spec1}
 \begin{align}
 \widetilde{p}(\x,s|\x_0)&=G(\x,s|\x_0)+\frac{1}{D}\sum_{n,m}{\mathcal V}_n(\x,s)\bigg [{\bm \Lambda}^{-1}_1(s){\bm \Lambda}_2(s)\bigg ]_{nm}{\mathcal V}^*_m(\x_0,s),\ \x \in \Omega \backslash \calU,\end{align}
 \begin{align} 
  \widetilde{q}(\x,s|\x_0)&=\frac{1}{D\kappa \alpha} \sum_{n,m}\overline{\mathcal V}_n(\x,s+\gamma)\bigg (\lambda_n(s)+\kappa(1-\alpha)\bigg)\bigg [{\bm \Lambda}^{-1}_1(s){\bm \Lambda}_2(s)\bigg ]_{nm}{\mathcal V}^*_m(\x_0,s)\nonumber\\
  &\quad -\frac{1}{D\kappa \alpha}  \sum_{n}\overline{\mathcal V}_n(\x,s+\gamma) {\mathcal V}^*_n(\x_0,s) , \quad  \x \in \calU.
 \end{align}
 \end{subequations}
 
There are two distinct challenges in using the spectral decompositions (\ref{spec1}) to determine the flux $\widetilde{J}(\x_0,s)$, and hence the FPT statistics, when $d\geq 2$:
\medskip

\noindent (i) Obtaining the eigenvalues and eigenfunctions of the D-to-N operators. One higher-dimensional example where the spectral decompositions of $\L_s$ and $\overline{\L}_s$ are known exactly is a partially absorbing sphere \cite{Grebenkov19b}. The rotational symmetry of $\calU$ means that if $\L_s$ and $\overline{\L}_s$ are expressed in spherical polar coordinates $(r,\theta,\phi)$, then the eigenfunctions are given by spherical harmonics, and are independent of the Laplace variable $s$ and the radius $r$:
\begin{equation}
v_{nm}(\theta,\phi)=\overline{v}_{nm}(\theta,\phi)=\frac{1}{R} Y_n^m(\theta,\phi),\quad n\geq 0, \ |m|\leq n.
\end{equation}
From orthogonality, it follows that the adjoint eigenfunctions are
\begin{equation}
v^*_{nm}(\theta,\phi)=\overline{v}_{nm}^*(\theta,\phi)=(-1)^m\frac{1}{R} Y_n^{-m}(\theta,\phi).
\end{equation}
(Note that eigenfunctions are labeled by the pair of indices $(nm)$.)
The corresponding eigenvalues are \cite{Grebenkov19b}
\begin{equation}
\lambda_n(s)=-\beta(s)\frac{k_n'(\beta(s) R)}{k_n(\beta(s) R)},\quad \overline{\lambda}_n(s)=\beta(s)\frac{i_n'(\beta(s) R)}{i_n(\beta(s) R)},
\end{equation}
where $\beta(s)=\sqrt{s/D}$, and $i_n(x),k_n(x)$ are modified spherical Bessel functions of the first and second kind, respectively.
Since the $n$th eigenvalue is independent of $m$, it has a multiplicity $2n+1$.
It is also possible to compute the projections of the boundary fluxes in (\ref{Vn}) by using appropriate series expansions of the corresponding Green's functions \cite{Grebenkov20}. 
\medskip

\noindent (ii) Numerically truncating the infinite series expansions in equations (\ref{spec1}) and inverting the matrix ${\bm \Lambda}_1(s)$.

\setcounter{equation}{0}

\section{Partially absorbing interval with a semipermeable barrier}

\begin{figure}[b!]
\centering
  \includegraphics[width=11cm]{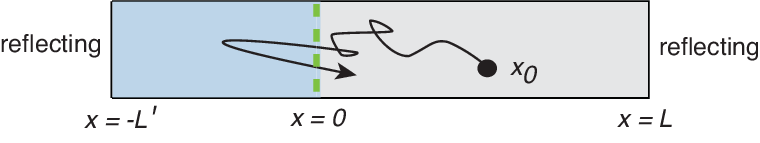}
  \caption{Partially absorbing substrate in 1D with $\calU=[-L',0]$ and $\Omega\backslash \calU=[0,L]$.}
  \label{fig2}
\end{figure}

In the case of a one-dimensional (1D) substrate, the D-to-N operators reduce to scalar multipliers so that the difficulties of higher dimensional interfaces are avoided. As a simple illustration of this,  consider a particle diffusing in the interval $\Omega=[-L',L]$ with a partially absorbing subinterval $\calU=[-L',0)$ and a semipermeable membrane at $x=0$, see Fig. \ref{fig2}. It follows that $\partial \Omega=\{-L',L\}$ and $\partial \calU=\{0\}$. The 1D version of equations (\ref{master}) takes the form
\begin{subequations} 
\label{1Dmaster}
\begin{eqnarray}
	\frac{\partial p(x,t|x_0)}{\partial t} &=& D\frac{\partial^2 p(x,t|x_0)}{\partial x^2}, \quad x\in (0,L),\quad \partial_x p(L,t|x_0)=0,,\\
	\frac{\partial q(x,t|x_0)}{\partial t} &=& D\frac{\partial^2 q(x,t|x_0)}{\partial x^2}-\gamma q(x,t|x_0),\quad x\in (-L',0),
	\end{eqnarray}
together with the
semipermeable boundary conditions
\begin{align}
 D\partial_xp(0^+,t|x_0)&= D\partial_x q(0^-,t|x_0) \equiv -  {\mathcal J}(x_0,t) \\
	{\mathcal J}(x_0,t)&=\kappa\bigg [(1-\alpha )p(0^+,t|x_0)- \alpha q(0^-,t|x_0)\bigg ].
\end{align}
\end{subequations}

The general solution (\ref{sool}) in Laplace space becomes
 \begin{subequations}
  \label{sool1D}
 \begin{align}
 \widetilde{p}(x,s|x_0)&= D \left . \partial_{x'} G(x',s|x)\right |_{x'=0}f(s)+G(x,s|x_0),\ x,x_0 \in (0,L],\\  \widetilde{q}(x,s|x_0)&=- D \left . \partial_{x'} \overline{G}(x',s+\gamma|x)\right |_{x'=0}\overline{f}(s),\ x\in [-L',0),
 \end{align}
 \end{subequations}
 for the unknown functions $f,\overline{f}$. Note that $\partial_{\sigma'}=-\partial_{x'}$.
 Equation (\ref{nabG}a) for the modified Helmholtz Green's function $G(x,s|x_0)$ reduces to
 \begin{subequations}
 \label{Go}
  \begin{align}
 &D\frac{d^2}{dx^2} G(x,s|x_0)-sG(x,s|x_0)=-\delta(x-x_0),\ 0<x,x_0 <L ,\\
 &G(0,s|x_0)=0,\ \left. \frac{d}{dx} G(L,s|x_0)\right |_{x=L}=0.
 \end{align}
 \end{subequations}
 The explicit solution is $G=G_L$ where
\begin{align}
 G_L(x, s| x_0) 
    &= \frac{\Theta(x_0 - x)g(x, s)\widehat{g}(x_0, s) +(x - x_0)g(x_0, s)\widehat{g}(x, s)}{\sqrt{sD}\cosh(\beta(s)L)},
\end{align}
$\beta(s)=\sqrt{s/D}$ and
\begin{eqnarray}
   g(x, s) = \sinh \beta(s) x,\quad \makebox{and} \quad \widehat{g}(x, s) =\cosh[\beta(s) (L-x)]. 
\end{eqnarray}
Similarly, $\overline{G}(x,s|x_0)=G_{L'}(-x,s|-x_0)$ for $x,x_0\in [-L',0)$. It also follows that
\begin{subequations}
\begin{align}
 D \left . \partial_{x'} G(x',s|x)\right |_{x'=0}&=\frac{\cosh(\beta(s)[L-x]}{\cosh(\beta(s)L)},\\ - D \left . \partial_{x'} \overline{G}(x',s+\gamma|x)\right |_{x'=0}&=\frac{\cosh(\beta(s+\gamma)[L'+x]}{\cosh(\beta(s)L')}.
 \end{align}
 \end{subequations}
 
Substituting for the Green's functions into equations (\ref{DtoN2}) and setting $\partial_{\sigma}=-\partial_x$ etc., we find that 
\begin{subequations}
\begin{align}
\L_s [f](s)&\equiv -Df(s)\left .\partial_{x}\partial_{x'} G(x',s|x)\right |_{x=x'=0}=f(s)\sqrt{\frac{s}{D}} \tanh(\sqrt{s/D}L) ,\\
\overline{\L}_s [f](s)&\equiv -Df(s)\left .\partial_{x}\partial_{x'} \overline{G}(x',s|x)\right |_{x=x'=0}= f(s)\beta(s)\tanh(\beta(s)L').
\end{align}
\end{subequations}
We deduce that for 1D diffusion, the D-to-N operators reduce to scalars with single eigenvalues $\lambda(s)=\beta(s) \tanh (\beta(s) L)$ and $\overline{\lambda}(s)=\beta(s)  \tanh (\beta(s) L')$. The unknown functions $f,\overline{f}$ are then determined from the 1D version of equations (\ref{fL0}):
\begin{subequations}
\label{fL}
\begin{align}
f(s)\beta(s) \tanh(\beta(s)L)-\partial_{x}G(0,s|x_0)&=-\overline{f}(s)\beta(s+\gamma) \tanh(\beta(s+\gamma)L'),\\
f(s)\beta(s) \tanh(\beta(s)L)-\partial_{x}G(0,s|x_0)&=-\frac{\kappa}{D}\bigg [(1-\alpha )f(s)- \alpha \overline{f}(s)\bigg ].
\end{align}
\end{subequations}
After some algebra, we find that
\begin{align}
\overline{f}(s)&=\Lambda(s)f(s),\quad 
f(s)=\frac{1}{D\Phi(s)}\frac{\cosh\beta(s) (L-x_0)}{\cosh\beta(s)L}
\end{align}
where
\begin{subequations}
\begin{align}
\Lambda(s,\gamma)&=\frac{1-\alpha}{\alpha}\frac{1}{1+\kappa^{-1}[D/\alpha]\beta(s+\gamma) \tanh(\beta(s+\gamma)L')},\\
\Phi(s,\gamma)&= \beta(s) \tanh(\beta(s)L)+\frac{1-\alpha}{\alpha}\frac{\beta(s+\gamma) \tanh(\beta(s+\gamma)L')}{1+\kappa^{-1}[D/\alpha]\beta(s+\gamma) \tanh(\beta(s+\gamma)L')} .
\end{align}
\end{subequations}

In the specific case $x_0=0^+$, the full solution has the particularly simple form
\begin{subequations}
\label{squid}
\begin{align}
\widetilde{p}(x,s|0^+)&=\frac{1}{\Phi(s,\gamma)D}\frac{\cosh\beta(s) (L-x)}{ \cosh\beta(s) L} ,\ x\in [0,L],\\
 \widetilde{q}(x,s|0^+)&=\frac{\Lambda(s,\gamma)}{\Phi(s,\gamma)D}\frac{\cosh\beta(s+\gamma) (L'+x)}{ \cosh\beta(s+\gamma) L'},\ x \in [-L',0],
\end{align}
\end{subequations}
From the 1D version of equation (\ref{TRobin}), the MFPT for absorption is
\begin{align}
T(0^+) = \frac{1}{\gamma} - \partial_s\widetilde{ {\mathcal J}}(0)  ,
\end{align}
with
\begin{align}
\widetilde{\mathcal J}(s)&=\kappa\bigg [(1-\alpha )\widetilde{p}(0,s|0^+)- \alpha \widetilde{q}(0,s|0^+)\bigg ]=\frac{\kappa}{D\Phi(s,\gamma)}\bigg [(1-\alpha )- \alpha \Lambda(s,\gamma)\bigg ]
\end{align}
Substituting for $\Phi(s,\gamma)$ and $\Lambda(s,\gamma)$ gives
\begin{align}
\widetilde{\mathcal J}(s)&=\frac{1-\alpha}{\alpha}\frac{\displaystyle 1}{\beta(s) \tanh(\beta(s)L)\left [\frac{1}{\displaystyle \beta(s+\gamma) \tanh(\beta(s+\gamma)L')}+\frac{\displaystyle D}{\displaystyle \alpha \kappa}\right ]+\frac{\displaystyle 1-\alpha}{\displaystyle  \alpha} }.
\end{align}
Hence, the MFPT for finite $\kappa$ is
\begin{align}
\label{TRobin2}
T (0^+) = \frac{\alpha }{1-\alpha}\frac{L}{\displaystyle \sqrt{\gamma D}\tanh(\sqrt{\gamma/D}L')}+\frac{\displaystyle L}{\displaystyle (1-\alpha) \kappa}+\frac{1}{\gamma}.
\end{align}
The final term on the right-hand side is the expected time for absorption when the particle is within the target, whereas the first two terms is the mean time spent outside the target, where no absorption can occur. The latter includes the contribution $L/(1-\alpha)\kappa$ associated with the effective ``resistance'' of the semi-permeable membrane to particle influx.

\setcounter{equation}{0}

\section{Encounter-based model of a partially absorbing target}

So far we have assumed that the absorption rate $\gamma$ is a constant. However, various absorption-based reactions are better modeled in terms of a reactivity that is a function of the amount of contact time between a particle and a target \cite{Bartholomew01,Filoche08}. 
That is, the substrate may need to be progressively
activated by repeated encounters with a diffusing
particle, or an initially highly reactive substrate may become less active due to multiple interactions with the particle (passivation). Recently, a so-called encounter-based approach has been developed for analyzing a more general class of partially absorbing surfaces $\partial \calU$ \cite{Grebenkov20,Grebenkov22}, and partially absorbing interiors $\calU$ with totally permeable interfaces \cite{Bressloff22,Bressloff22a}. Here we extend the latter to the case of a semipermeable interface $\partial \calU$.

\subsection{Occupation time propagator}

The basic idea of the encounter-based approach is to consider the joint probability density or generalized propagator $P(\x,a,t)$ for the pair $(\X_t,\calA_t)$, where $\calA_t$ is a Brownian functional that specifies the amount of contact time with the target over the time interval $[0,t]$ (in the absence of absorption). In the case of an interior target or trap $\calU$, the functional $\calA_t$ is identified with the occupation time \cite{Majumdar05}
\begin{equation}
\label{occ}
\calA_t=\int_{0}^tI_{\calU}(\X_{\tau})d\tau .
\end{equation}
Here $I_{\calU}(\x)$ denotes the indicator function of the set $\calU\subset \Omega$, that is, $I_{\calU}(\x)=1$ if $\x\in \calU$ and is zero otherwise. The effects of partial absorption are then incorporated 
 by introducing the stopping time 
$
{\mathcal T}=\inf\{t>0:\ \calA_t >\widehat{\calA}\}$,
 with $\widehat{\calA}$ a randomly distributed occupation time threshold. Given the probability distribution $\Psi(a) = \P[\widehat{\calA}>a]$, the marginal probability density for particle position is defined according to
\[p^{\Psi}(\x,t)d\x=\P[\X_t \in (\x,\x+d\x), \ t < {\mathcal T}].\]
Since $\calA_t$ is a nondecreasing process, the condition $t < {\mathcal T}$ is equivalent to the condition $\calA_t <\widehat{\calA}$. This implies that 
\begin{align*}
p^{\Psi}(\x,t)d\x&=\P[\X_t \in (\x,\x+d\x), \ \calA_t < \widehat{\calA}]\\
&=\int_0^{\infty} da \ \psi(a)\P[\X_t \in (\x,\x+d\x), \ \calA_t < a]\\
&=\int_0^{\infty} da \ \psi(a)\int_0^{a} da' [P(\x,a',t)d\x],
\end{align*}
where $\psi(a)=-\Psi'(a)$ and $P(\x,a,t)$ denotes the joint probability density for the pair $(\X_t,\calA_t)$. 
Using the identity
\[\int_0^{\infty}dv\ f(v)\int_0^{v} dv' \ g(v')=\int_0^{\infty}dv' \ g(v')\int_{v'}^{\infty} dv\ f(v)\]
for arbitrary integrable functions $f,g$, it follows that
\begin{equation}
\label{bob}
p^{\Psi}(\x,t)=\int_0^{\infty} \Psi(a)P(\x,a,t)da.
\end{equation}
  
Let $P(\x,a,t|\x_0)$ denote the occupation time propagator under the initial conditions $\X_0=\x_0$ and $\calA_0=0$.
It follows that
 \begin{eqnarray}
 \label{A1}
& P(\x,a,t|\x_0)=\bigg \langle \delta\left (a -\calA_t \right )\bigg \rangle_{\X_0=\x_0}^{\X_t=\x} ,
 \end{eqnarray}
 where expectation is taken with respect to all random paths realized by $\X_{\tau}$ between $\X_0=\x_0$ and $\X_t=\x$. 
 Using the Feynman-Kac formula, it can be shown that away from the boundaries $\partial \Omega $ and $\partial \calU^{\pm}$, the propagator satisfies a BVP of the form \cite{Bressloff22a}
 \begin{subequations}
\label{calP}
\begin{align}
&\frac{\partial P(\x,a,t|\x_0)}{\partial t}=D\nabla^2 P(\x,a,t|\x_0)-I_{\calU}(\x) \frac{\partial P}{\partial a}(\x,a,t|\x_0) \ - \delta(a)I_{\calU}(\x) P(\x,0,t|\x_0), 
\end{align}
\end{subequations}
for all $\x \in\Omega$.
 Since the reflecting boundary condition on $\partial \Omega$ and the semipermeable boundary conditions across $\partial \calU$ are independent of the occupation time, they also hold for the propagator. We thus have the following BVP for the occupation time propagator:
\begin{subequations}
\label{Pocc}
\begin{align}
 \frac{\partial P(\x,a,t|\x_0)}{\partial t}&=D\nabla^2 P(\x,a,t|\x_0), \ \x \in \Omega\backslash \calU,\\
 \nabla P(\x,a,t|\x_0) \cdot \n&=0, \  \x \in \partial \Omega, \\
 \frac{\partial Q(\x,a,t|\x_0)}{\partial t}+\frac{\partial Q(\x,a,t|\x_0)}{\partial a}&=D\nabla^2 Q(\x,a,t|\x_0)  -\delta(a)Q(\x,0,t|\x_0) 
\end{align}
for $ \x \in \calU$, and
\begin{align}
 D\nabla P(\x,a,t|\x_0)\cdot \n_0&= D\nabla Q(\x,a,t|\x_0)\cdot \n_0\equiv -  {\mathcal J}(\x,a,t|\x_0) \\
	{\mathcal J}(\x,a,t|\x_0)&=\kappa\bigg [(1-\alpha )P(\x,a,t|\x_0)- \alpha Q(\x,a,t|\x_0)\bigg ] ,\quad \x   \in \partial \calU .
\end{align}
	\end{subequations}
	We now denote the propagator within the target $\calU$ by $Q$.
The initial conditions are $P(\x,a,0|\x_0)=\delta(\x-\x_0)\delta(a)$, $Q(\x,a,0|\x_0)=0$, assuming that the particle starts in the non-absorbing region. (The analysis is easily modified if $\x_0\in \calU$.) 
One interesting observation is that equation (\ref{Pocc}c) takes the form of an age-structured model, reflecting the fact that whenever the particle is within the target interior $\calU$, the occupation time $\calA_{t}$ increases at the same rate as the absolute time $t$. (Age-structured models are typically found within the context of birth-death processes in ecology and cell biology, where the birth and death rates of individual organisms and cells depend on their age
\cite{McKendrick25,Foerster59,Iannelli17}.) Finally, note that the term involving the Dirac delta function $\delta(a)$ in equation (\ref{Pocc}c) ensures that the probability of being in the boundary layer is zero if the occupation time is zero.

\subsection{Marginal probability density and flux}
Laplace transforming equations (\ref{Pocc}a,c) with respect to $a$ and setting
\begin{align}
\widetilde{P}(\x,z,t|\x_0)&=\int_0^{\infty} \e^{-za} P(\x,a,t|\x_0)da,\quad \widetilde{Q}(\x,z,t|\x_0)=\int_0^{\infty} \e^{-za} Q(\x,a,t|\x_0) da,
\end{align}
yields
\begin{subequations}
\label{PoccLT}
\begin{align}
 \frac{\partial \widetilde{P}(\x,z,t|\x_0)}{\partial t}&=D\nabla^2 \widetilde{P}(\x,z,t|\x_0) , \ \x \in \Omega\backslash \calU,\\ \frac{\partial \widetilde{Q}(\x,z,t|\x_0)}{\partial t}&=D\nabla^2 \Q(\x,z,t|\x_0) -z \Q(\x,z,t|\x_0),\  \x \in \calU,
 \end{align}
	\end{subequations}
	together with the Laplace transformed versions of the boundary conditions (\ref{Pocc}b,d,e).
We thus recover the BVP (\ref{master}) for diffusion in a domain with a partially absorbing trap $\calU$ with a constant rate of absorption $z$. This establishes that the original BVP for a partially absorbing trap with a constant absorption rate $z$ is recovered by taking the occupation time threshold to be an exponential random variable. That is, $\Psi(a)=\e^{-z a}$. In other words, the marginal density $p(\x,t|\x_0)$ for a constant absorption rate $z$ is equivalent to the Laplace transform $\widetilde{P}(\x,z,t|\x_0)$ of the occupation time propagator. Assuming that the inverse Laplace transform exists, we have the general result
 \begin{equation}
 \label{pca2}
 p^{\Psi}(\x,t)=\int_0^{\infty} \Psi(a)\, \mbox{LT}^{-1}[ \widetilde{P}](\x,a,t)]da.
\end{equation}
 
\begin{figure}[t!]
\includegraphics[width=7cm]{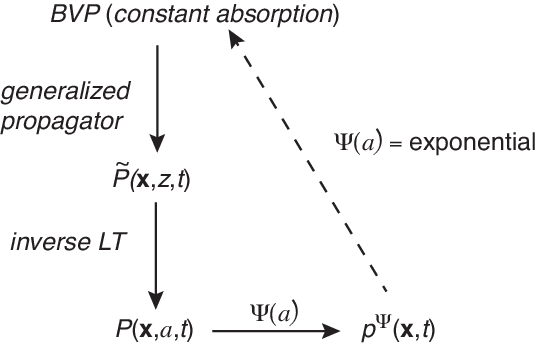}
\caption{Diagram illustrating the encounter-based framework for diffusion in a domain with a partially absorbing target. The solution of the BVP in the case of a constant absorption rate $\gamma$ generates the Laplace transform $\widetilde{P}(\x,z,t)$ of the occupation time propagator $P(\x,a,t)$. The inverse LT determines the marginal probability density $p^{\Psi}(\x,t)$ according to equation (\ref{pca2})}
\label{fig3}
\end{figure}

 The general probabilistic framework for analyzing single-particle diffusion in partially absorbing media is summarized in the commutative diagram of Fig. \ref{fig3}. One of the challenges of implementing the encounter-based method is that solutions of the classical BVP with a constant rate of absorption $z$ tend to have a non-trivial parametric dependence on the Laplace variable $z$, which makes it difficult to calculate the inverse transform. This is clear from the spectral decomposition given by equations (\ref{spec1}) on replacing $\gamma$ by the Laplace variable $z$. (In contrast, solving the Robin BVP for a reactive surface in terms of the spectrum of an associated D-to-N operator yields a series expansion that is easily inverted with respect to the Laplace variable $z$ conjugate to the boundary contact time or local time \cite{Grebenkov20,Grebenkov22}.) 
In order to invert the $z$-Laplace transforms term by term in equations (\ref{spec1}), we require these infinite series to be uniformly convergent. Assuming that this is the case, one then has to determine how many terms in the series are required in order to obtain a given level of accuracy for quantities of interest such as the MFPT. After taking the $s\rightarrow 0$ limit, accuracy will depend on the choice of the stopping time distribution $\Psi$. That is, although the spectral decomposition of the propagator is independent of $\Psi$, the numerical truncation of the corresponding expansion of the MFPT will be $\Psi$-dependent. 

Finally, note that the statistics of the FPT density for non-exponential $\Psi$ proceeds along analogous lines to the exponential case. In particular, the FPT moment generator is given by the Laplace transform of the target flux $J^{\Psi}(\x,_0,t)$. The generalized survival probability is
\begin{equation}
\label{Socc}
S^{\Psi}(\x_0,t)= \int_{\Omega\backslash \calU}  p^{\Psi}(\x,t|\x_0)d\x+\int_{ \calU}  q^{\Psi}(\x,t|\x_0)d\x,
\end{equation}
with
\begin{equation}
\label{bob2}
p^{\Psi}(\x,t)=\int_0^{\infty} \Psi(a)P(\x,a,t)da,\quad q^{\Psi}(\x,t)=\int_0^{\infty} \Psi(a)Q(\x,a,t)da.
\end{equation}
Differentiating with respect to $t$ and using equations (\ref{Pocc}a) and (\ref{Pocc}c) gives
\begin{eqnarray}
\frac{\partial S^{\Psi}(\x_0,t)}{\partial t}&=&D\int_{\Omega\backslash \calU} \nabla^2 p^{\Psi}(\x,t|\x_0)d\x+D \int_{  \calU}\nabla^2 q^{\Psi}(\x,t|\x_0)d\x\nonumber \\
&&\quad -\int_{\calU} \int_0^{\infty} \psi(a) Q (\x,a,t|\x_0)da\ d\x.
\label{Qocc}
\end{eqnarray}
Applying the divergence theorem to the first two integrals on the right-hand side, imposing the Neumann boundary condition on $\partial \Omega$ and flux continuity at $\partial \calU$ shows that these two integrals cancel. The result is then
\begin{eqnarray}
\frac{\partial S^{\Psi}(\x_0,t)}{\partial t}=-\int_{\calU} \int_0^{\infty} \psi(a) Q(\x,a,t|\x_0)da\ d\x \equiv -J^{\Psi}(\x_0,t).
\label{flum}
\end{eqnarray}
In the exponential case, $\psi(a)=\gamma \e^{-\gamma a}$, we recover equation (\ref{Qin2}).

\subsection{One-dimensional substrate}

Let us return to the 1D example of section 4. Equation (\ref{squid}b) implies that
\begin{align}
\label{aboo}
\widetilde{Q}(x,z,s|0^+)&=\frac{\Lambda(s,z)}{\Phi(s,z)D}\frac{\cosh\beta(s+z) (L'+x)}{ \cosh\beta(s+z) L'},\ x \in [-L',0],
\end{align}
In order to determine the flux $ \widetilde{J}^{\Psi}(0^+,s)$ given by Laplace transforming the 1D version of equation (\ref{flum}) with respect to $t$, we need to calculate the inverse Laplace transform of equation (\ref{aboo}) with respect to $z$. This is relatively straightforward in the limit 
 $L'\rightarrow \infty$, since $\tanh(\sqrt{(s+z)/D}L')\rightarrow 1$ and
  \begin{align}
\widetilde{Q}(x,z,s|0^+)&=\frac{[(1-\alpha)/\alpha D]\e^{\beta(s+z)  x}}{\beta(s) \tanh(\beta(s)L)\left [1+\frac{\dis D\beta(s+z)}{\dis \kappa \alpha}\right ]+\frac{\dis 1-\alpha}{\dis \alpha} \beta(s+z) } 
\end{align}
for $\ x \in (-\infty,0].$ Integrating with respect to $x$ then gives
\begin{align}
\int_{-\infty}^0 \widetilde{Q}(x,z,s|0^+)dx&=\frac{[(1-\alpha)/\alpha ]}{D\beta(s) \tanh(\beta(s)L)\left [\beta(s+z)+\frac{\dis  s+z}{\dis \kappa \alpha}\right ]+\frac{\dis 1-\alpha}{\dis \alpha}  (s+z) } .
\label{Qoo}
\end{align}
We thus find that
\begin{align}
T(0^+)&=-\partial_s\widetilde{J}^{\Psi}(0^+,0)=-\int_0^{\infty}da\, \psi(a)  \mbox{LT}^{-1}\bigg [\int_{-\infty}^0 \partial_s\widetilde{Q}(x,z,0|0^+)dx\bigg ].
\end{align}
Differentiating equation (\ref{Qoo}) with respect to $s$, we find that
\begin{align}
T(0^+)&=\int_0^{\infty}da\, \psi(a)  \mbox{LT}^{-1}\bigg [\frac{1}{z^2}\bigg (
1+\frac{\alpha L}{1-\alpha}\left (\sqrt{\frac{z}{D}}+\frac{z}{\kappa \alpha }\right ) \bigg )\bigg]\nonumber \\
&=\int_0^{\infty}da\, \psi(a)  \bigg [a+\frac{\alpha L}{1-\alpha}\left (2\sqrt{\frac{a}{\pi D}}+\frac{1}{\kappa \alpha}\right )\bigg ]\nonumber \\
&=\E[a]+\frac{2\alpha L\E[\sqrt{a}]}{[1-\alpha]\sqrt{\pi D}}+\frac{L}{[1-\alpha]\kappa}.
\label{cool2}
\end{align}

A few comments are in order. First, in the case of the exponential density $\psi(a)=\gamma \e^{-\gamma a}$, we have $\E[a] =\gamma^{-1}$ and 
$\E[\sqrt{a}]=\sqrt{\pi/\gamma}/2$.
Hence, equation (\ref{cool2}) reduces to equation (\ref{TRobin2}) in the limit $L'\rightarrow \infty$. Second, the $\kappa$-dependent term is independent of the occupation time distribution $\Psi$. Finally, in the case of a non-exponential distribution $\Psi$, the MFPT only exists if the corresponding moments $\E[a]$ and $\E[\sqrt{a}]$ are finite.

\setcounter{equation}{0}

\section{Snapping out Brownian motion for semipermeable interfaces}

\begin{figure}[b!]
\includegraphics[width=11cm]{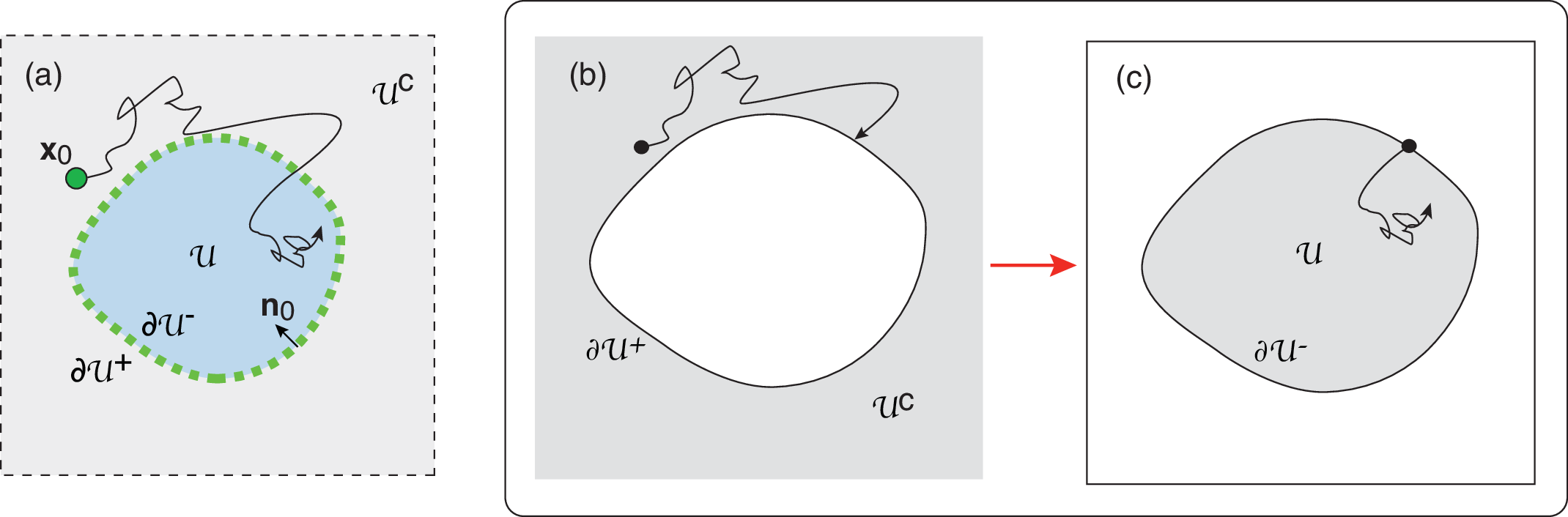}
\caption{Decomposition of (a) snapping out BM into two partially reflected BMs corresponding to (b) $\X_t\in \calU^c$ and (c) $\X_t\in \calU$ , respectively.}
\label{fig4}
\end{figure}

The encounter-based framework for absorbing targets can also be used to develop a more general probabilistic model of single-particle diffusion through semi-permeable interfaces, by combining it with so-called snapping out Brownian motion (BM) \cite{Bressloff22b,Bressloff23,Bressloff23b}. The latter was originally formulated for 1D single-particle diffusion through a semipermeable barrier \cite{Lejay16,Lejay18,Brobowski21}, but has recently been extended to higher spatial dimensions \cite{Bressloff23}. In order to present the basic theory, we ignore the effects of absorption by setting $\gamma=0$. Snapping out BM sews together successive rounds of partially reflecting BM that are restricted to either the interior or the exterior of $\partial \calU$, see Fig. \ref{fig4}. 
Suppose that the particle starts in the domain $ {\calU^c}=\Omega\backslash \calU$ ($\calU$). It realizes reflected BM until it is killed when its local time on $\partial \calU^+$ ($\partial \calU^-$) is greater than an exponentially distributed random threshold. (This is analogous to the killing of BM when the occupation time $\calA_t$ spent within a target  interior $\calU$ exceeds an exponentially distributed random threshold, signifying an absorption event, see section 5.) Let $\y\in \partial \calU$ denote the point on the boundary where killing occurs. The stochastic process immediately restarts as a new round of partially reflected BM, either from $\y^+$ into $\calU^c$ or from $\y^-$ into $\calU$. These two possibilities occur with the probabilities $\alpha$ and $1-\alpha$, respectively. Subsequent rounds of partially reflected BM are generated in the same way. We thus have a stochastic process on the set ${\mathbb G}=\overline{\calU}\cup \overline{\calU^c}$. It can be proven that the probability density of sample paths generated by snapping out BM evolves according to equations (\ref{master}a) and (\ref{master}b) for $\gamma=0$, together with the semipermeable boundary conditions (\ref{master}c) and (\ref{master}d) \cite{Lejay16,Bressloff22b,Bressloff23}. (One version of the proof for the 1D case is given in section 6.3.) For simplicity, we set $\alpha=1/2$ in the following.

\subsection{Partially reflected BMs in $\calU$ and $\calU^c$}

Consider a Brownian particle diffusing in the bounded domain $ \calU^c$, see Fig. \ref{fig4}(b) with $\partial \calU^+$ totally reflecting. Let $\X_t$ denote the position of the particle at time $t$. In order to write down a stochastic differential equation (SDE) for $\X_t$, we introduce the boundary local time \cite{Levy40,Ito63,Dynkin65,McKean75,Majumdar05}
\begin{equation}
\label{loc}
\ell_t^+=\lim_{\epsilon\rightarrow 0} \frac{D}{\epsilon} \int_0^t\Theta(\epsilon-\mbox{dist}(\X_{\tau},\partial \calU^+))d\tau,
\end{equation}
where $\Theta$ is the Heaviside function, and $\mbox{dist}(\X_{\tau},\partial \calU^+)$ denotes the shortest Euclidean distance of $X_{\tau}$ from the boundary $\partial \calU^+$.
The corresponding SDE takes the form 
\begin{equation}
d\X_t =\sqrt{2D}d{\bf W}_t-\n_0(\X_t)  d\ell_t^+,
\end{equation}
where ${\bf W}_t$ is a $d$-dimensional Brownian motion and $\n_0(\X_t)$ is the inward unit normal at the point $\X_t \in \partial \calU$. The differential $d\ell^+_t$ can be expressed in terms of a Dirac delta function:
\begin{equation}
d\ell^+_t=Ddt\left (\int_{\partial \calU^+}\delta(\X_t-\y)d\y\right ) .
\end{equation}
{Partially reflected BM in} $\calU$ is then obtained by stopping the stochastic process $\X_t$ when the local time $\ell_t^+$ exceeds a random exponentially distributed threshold $\widehat{\ell}^+$ \cite{Grebenkov20}. That is, the particle is absorbed somewhere on $\partial \calU^+$ at the stopping time
 \begin{equation}
\label{exp}
{\mathcal T}^+=\inf\{t>0:\ \ell_t^+ >\widehat{\ell}^+\},\quad \P[\widehat{\ell}^+>\ell]  =\e^{-\kappa_0 \ell/D}.
\end{equation}

Consider the local time propagator $P(\x,\ell,t|\x_0)$ for the pair $(\X_t,\ell_t)$, which evolves according to \cite{Grebenkov20}
\begin{subequations} 
\label{Ploc}
\begin{align}
 &\frac{\partial P(\x,\ell,t|\x_0)}{\partial t}=D\nabla^2 P(\x,\ell,t|\x_0),\ \x \in \Omega\backslash \calU, \ \nabla P(\x,\ell,t|\x_0) \cdot \n =0, \  \x\in \partial \calU^c,\\
 &-D\nabla P(\x,\ell,t|\x_0) \cdot \n_0= D P(\x,\ell=0,t|\x_0) \ \delta(\ell)  +D\frac{\partial}{\partial \ell} P(\x,\ell,t|\x_0),  \x\in \partial \calU.
\end{align}
\end{subequations}
This can be derived using a Feynman-Kac formula along analogous lines to the occupation time propagator of section 5, see Ref. \cite{Bressloff22a}.
Equations (\ref{Ploc}) are supplemented by the initial condition $P(\x,\ell,0|\x_0)=\delta(\x-\x_0)\delta(\ell)$.
Laplace transforming the local time BVP (\ref{Ploc}) with respect to $\ell$ and setting
\begin{equation}
\widetilde{P}(\x,\omega,t|\x_0)=\int_0^{\infty}\e^{-\omega \ell}P(\x,\ell,t|\x_0)d\ell
\end{equation}
yields
\begin{subequations}
\label{PlocLT0}
\begin{align}
 &\frac{\partial \widetilde{P}(\x,\omega,t|\x_0)}{\partial t}=D\nabla^2 \widetilde{P}(\x,\omega,t|\x_0),\ \x \in \Omega\backslash \calU, \ \nabla \widetilde{P}(\x,\omega,t|\x_0) \cdot \n =0,\ \x\in \partial \Omega,\\
 &-\nabla \widetilde{P}(\x,\omega,t|\x_0) \cdot \n_0= \omega \PP(\x,\omega,t|\x_0),\  \x\in \partial \calU^+,
\end{align}
\end{subequations}
and $\widetilde{P}(\x,\omega,0|\x_0)=\delta(\x-\x_0)$.
We see that equation (\ref{PlocLT0}b) is a classical Robin boundary condition on $\partial \calU^+$ with a constant reactivity $\kappa_0=\omega D$. Hence, the Robin boundary condition is equivalent to an exponential law for the local time threshold $\widehat{\ell}^+$. Following Ref. \cite{Grebenkov20}, we now modify the rule for killing each round of partially reflected BM by taking $\widehat{\ell}^{+}$ to have a non-exponential distribution $\Psi^{+}(\ell)$. We then define the corresponding marginal probability density according to
\begin{align}
\label{0pjay}
p_+^{\Psi}(\x,t|\x_0) &=   \int_0^{\infty} \Psi^+(\ell)P(\x,\ell,t|\x_0)d\ell ,\quad \x\in \calU^c.
\end{align} 
Multiplying both sides of the boundary condition (\ref{Ploc}b) by $\Psi(\ell)$ and integrating by parts with respect to $\ell$ shows that
\begin{eqnarray}
j_+^{\Psi}(\x,t|\x_0)\equiv -D\nabla p_+^{\Psi}(\x,t|\x_0) \cdot \n_0=D\int_0^{\infty}\psi(\ell) P(\x,\ell,t|\x_0)d\ell, \  \x\in \partial \calU^+,
\label{jj}
\end{eqnarray}
with $\psi^+(\ell)=-{\Psi^+}'(\ell)$.
We have used equation (\ref{Ploc}b) and the identity $\Psi^+(0)=1$. Integrating with respect to points on the boundary then gives
\begin{equation}
\label{J}
{J}_+^{\Psi}(\x_0,s)=\int_{\partial \calU}j_+^{\Psi}(\x,t|\x_0)d\sigma\equiv D \int_{\partial \calU}\left [\int_0^{\infty}\psi^+(\ell){P}(\x,\ell,t|\x_0)d\ell \right ]d\sigma.
\end{equation}
Finally, we can determine ${P}(\x,\ell,t|\x_0) $ by inverting the solution $\widetilde{P}(\x,\omega,t|\x_0)$ to the Robin BVP with respect to $\omega$, which is the local time analog of equation (\ref{pca2}). In other words, a commutative diagram of the form shown in Fig. \ref{fig3} also applies to the local time propagator.

An analogous construction holds for partially reflected BM in $ \calU$, see Fig. \ref{fig4}(c). Given the local time
\begin{equation}
\label{locp}
\ell_t^-=\lim_{\epsilon\rightarrow 0} \frac{D}{\epsilon} \int_0^tH(\epsilon-\mbox{dist}(\X_{\tau},\partial \calU^-))d\tau,
\end{equation}
and stopping time
\begin{equation}
\label{expp}
{\mathcal T}^-=\inf\{t>0:\ \ell_t^- >\widehat{\ell}^-\},\quad \P[\widehat{\ell}^->\ell] =\e^{-\kappa_0 \ell/D},
\end{equation}
we introduce the local time propagator $Q(\x,\ell,t|\x_0)$, which evolves according to
\begin{subequations} 
\label{Qloc}
\begin{align}
 &\frac{\partial Q(\x,\ell,t|\x_0)}{\partial t}=D\nabla^2 Q(\x,\ell,t|\x_0),\ \x \in   \calU, \\
 &D\nabla Q(\x,\ell,t|\x_0) \cdot \n_0= D Q(\x,\ell=0,t|\x_0) \ \delta(\ell)  +D\frac{\partial}{\partial \ell} Q(\x,\ell,t|\x_0),  \x\in \partial \calU.
\end{align}
\end{subequations}
Laplace transforming with respect to $\ell$ yields the following Robin BVP:
\begin{subequations}
\label{QlocLT0}
\begin{align}
 &\frac{\partial \widetilde{Q}(\x,\omega,t|\x_0)}{\partial t}=D\nabla^2 \widetilde{Q}(\x,\omega,t|\x_0),\ \x \in   \calU,\\
 &\nabla \widetilde{Q}(\x,\omega,t|\x_0) \cdot \n_0= \omega \widetilde{Q}(\x,\omega,t|\x_0),\  \x\in \partial \calU,
\end{align}
\end{subequations}
and $\widetilde{Q}(\x,\omega,0|\x_0)=\delta(\x-\x_0)$ assuming that $\x_0\in \calU$. Finally, given a local time threshold distribution $\Psi^-(\ell)$, the generalized marginal density is
\begin{align}
p_-^{\Psi}(\x,t|\x_0) &=  
  \int_0^{\infty}\Psi^-(\ell)Q(\x,\ell,t|\x_0)d\ell,\quad \x \in \calU.
\end{align}

\subsection{Renewal equations for snapping out BM}

The crucial step in formulating snapping out BM is sewing together successive rounds of partially reflected BM. As we have recently shown \cite{Bressloff22,Bressloff23}, this can be achieved by constructing renewal equations that relate the full probability density $\rho^{\Psi}$ of snapping out BM to the corresponding probability densities $p_{\pm}^{\Psi}$. First, it is convenient to consider a distribution of  initial conditions by setting
\begin{subequations}
\label{pjay}
\begin{align}
p_+^{\Psi}(\x,t) &=\int_{\calU^c} \left [\int_0^{\infty} \Psi^+(\ell)P(\x,\ell,t|\x_0)d\ell \right ]h(\x_0)d\x_0,\quad \x\in \calU^c,\\
 p_-^{\Psi}(\y,t)&=\int_{\calU} \left [\int_0^{\infty}\Psi^-(\ell)Q(\y,\ell,t|\y_0)d\ell\right ]\overline{h}(\y_0)d\y_0,\quad \y \in \calU,
\end{align} 
\end{subequations}
with
\begin{equation}
\int_{\calU^c}h(\x_0)d\x_0+\int_{\calU} \overline{h}(\y_0)d\y_0=1.
\end{equation}
Denote the probability density of generalized snapping out BM given $\x_0={\mathbb G}$ by $\rho^{\Psi}(\x,t|\x_0)$
and set
\begin{equation}
\label{rhojay}
\rho^{\Psi}(\x,t)=\int_{\calU^c} \rho^{\Psi}(\x,t|\x_0)h(\x_0)d\x_0+\int_{\calU} \rho^{\Psi}(\x,t|\y_0)\overline{h}(\y_0)d\y_0.
\end{equation}
It can then be shown that $\rho^{\Psi}(\x,t)$ satisfies the first renewal equation \cite{Bressloff23}
 \begin{align}
  \label{Frenewal}
  \rho^{\Psi}(\x,t)&=I_{\calU^c}(\x)p_+^{\Psi}(\x,t)+I_{\calU}(\x)p_-^{\Psi}(\x,t) +\frac{\kappa}{2}\int_0^t  d\tau\int_{\partial \calU}d\z \\
  &\quad \times \bigg [\rho^{\Psi}(\x,t-\tau |\z^+) +\rho^{\Psi}(\x,t-\tau |\z^-) \bigg ] f^{\Psi}(\z,\tau)\nonumber
   \end{align}
for $ \x\in {\mathbb G}$. In addition, $f^{\Psi}(\z,\tau)$ is the FPT density for the particle to be killed at time $\tau$ and a point $\z\in \calU$ for the given distribution of initial conditions. That is,
\begin{align}
\label{ff}
f^{\Psi}(\z,\tau)=\int_{\calU^c} j_+^{\Psi}(\z^+,t|\x_0)h(\x_0)d\x_0+\int_{\calU} j_-^{\Psi}(\z^-,t|\y_0)\overline{h}(\y_0) d\y_0.
\end{align}
The first two terms on the right-hand side of equation (\ref{Frenewal}) represent all sample trajectories that have never been absorbed by the boundaries $\partial \calU^+$ and $\partial \calU^-$, respectively. The integrand for a given $\z\in \partial \calU$ represents all trajectories that were first absorbed (stopped) at time $\tau$ and position $\z$, and then switched to either the domain $ {\calU^c}$ or $\calU$ with probability 1/2, after which multiple killing events can occur before reaching $\x$ at time $t$. The probability that the first stopping event occurred at $\z$ in the interval $(\tau,\tau+d\tau)$ is $ f^{\Psi}(\z,\tau)d\tau$. Finally, it is necessary to integrate with respect to all first stopping positions $\z$. 

Laplace transforming the renewal equation (\ref{Frenewal}) with respect to time $t$ and using the convolution theorem gives
 \begin{align}
  \widetilde{\rho}^{\Psi}(\x,s)&=I_{\calU^c}(\x)\widetilde{p}_+^{\Psi}(\x,s)+I_{\calU}(\x)\widetilde{p}_-^{\Psi}(\x,s) +\frac{\kappa}{2} \int_{\partial \calU}d\z \\
  &\quad \times \bigg [ \widetilde{\rho}^{\Psi}(\x,s |\z^+) + \widetilde{\rho}^{\Psi}(\x,s |\z^-) \bigg ] \widetilde{f}^{\Psi}(\z,s) .\nonumber
   \end{align}
In order to determine the factor $\widetilde{\rho}^{\Psi}(\x,s |\z^+)+\widetilde{\rho}^{\Psi}(\x,s |\z^-)$ we set $g(\x_0)=\delta(\x_0-\z^+) $ and $\overline{g}(\y_0)=\delta(\y_0-\z^-) $ in equations (\ref{pjay}), (\ref{rhojay}) and (\ref{ff}). This gives
\begin{align}
  \label{Frenewal2}
  &\widetilde{\rho}^{\Psi}(\x,s|\z^+)+\widetilde{\rho}^{\Psi}(\x,s|\z^-)=I_{\calU^c}(\x)\widetilde{p}_+^{\Psi}(\x,s|\z^+)+I_{\calU}(\x)\widetilde{p}_-^{\Psi}(\x,s|\z^-) \nonumber\\
  &\quad +\frac{\kappa}{2} \int_{\partial \calU}d\z \bigg [ \widetilde{\rho}^{\Psi}(\x,s |\z^+) + \widetilde{\rho}^{\Psi}(\x,s |\z^-) \bigg ] \bigg [ \widetilde{j}_+^{\Psi}(\z^+,s|\z^+)+\widetilde{j}_-^{\Psi}(\z^-,s|\z^-)\bigg ] .
   \end{align}
   For general geometries, solving this implicit integral equation for $\widetilde{\rho}^{\Psi}(\x,s|\z^+)+\widetilde{\rho}^{\Psi}(\x,s|\z^-)$ is nontrivial. Therefore, we will illustrate the theory using a 1D interface.
   
   \subsection{Snapping out BM in an interval}
   
   Let us return to the 1D example shown in Fig. \ref{fig2} with zero absorption ($\gamma=0$). After Laplace transforming with respect to $t$, the 1D version of equations (\ref{PlocLT0}) becomes
  \begin{subequations}
\label{RobinLT}
\begin{align}
&D\frac{\partial^2\widetilde{P}(x,\omega,s|x_0)}{\partial x^2}-s \widetilde{P}(x,\omega,s|x_0)=-\delta(x-x_0),\\
&\partial_x\widetilde{P}(0,\omega,s|x_0)=\omega \widetilde{P}(0,\omega,s|x_0),\quad \partial_x\widetilde{P}(L,\omega,s|x_0)=0,
\end{align}
\end{subequations}
with $0<x,x_0<L$.
We can identify $\widetilde{P}(x,\omega,s|x_0)$ as a Green's function of the modified Helmholtz equation on $[0,L]$, similar to $G$ of equation (\ref{Go}) but with different boundary conditions
 \begin{align}
 \widetilde{P}(x,\omega,s|x_0)= \left \{ \begin{array}{cc} A_L(\omega,s) g(x,\omega,s)\widehat{g}_L(x_0,\omega ,s), & 0\leq x\leq x_0\\ & \\
 A_L(\omega,s)g(x_0,\omega,s)\widehat{g}(x,\omega,s), & x_0\leq x\leq L\end{array}
 \right .
 \label{solVN}
 \end{align}
 with
 \begin{align}
g(x,\omega,s)
&=\frac{\sqrt{sD}\cosh(\beta(s)x)+\omega D\sinh(\beta(s)x)}{\sqrt{sD}+\omega D},\ \widehat{g}_L(x,s)= \cosh(\beta(s)(L-x))
\end{align}
and
 \begin{equation}
  A_L(\omega,s)= \frac{1}{\sqrt{sD}}\frac{\sqrt{sD}+\omega D} {\sqrt{sD}\sinh(\beta(s)L)+\omega D\cosh(\beta(s)L)}.
  \end{equation}
 Similarly, $\widetilde{Q}(x,\omega,s|x_0)$ satisfies equations (\ref{RobinLT}) for $L\rightarrow L'$ and $x,x_0\rightarrow -x,-x_0$. Hence,
 \begin{align}
 \widetilde{Q}(x,\omega,s|x_0)= \left \{ \begin{array}{cc}A_{L'}(\omega,s) g(-x,\omega,s)\widehat{g}_{L'}(-x_0,\omega ,s), & x_0\leq x \leq 0\\ & \\
 A_{L'}(\omega,s)g(-x_0,\omega,s)\widehat{g}_{L'}(-x,\omega ,s), & x\leq x_0 \leq 0\end{array}
 \right .
 \label{solVNq}
 \end{align}

 The Laplace transformed propagators have simple poles in the complex $\omega$-plane and can thus be inverted straightforwardly. For the sake of illustration, suppose that $x_0=0$. Then
 \begin{subequations}
 \begin{align}
 \widetilde{P}(x,\omega,s|0)&=\frac{\cosh(\beta(s)(L-x))} {\sqrt{sD}\sinh(\beta(s)L)+\omega D\cosh(\beta(s)L)},\\ \widetilde{Q}(x,\omega,s|0)&=\frac{\cosh(\beta(s)(L'+x))} {\sqrt{sD}\sinh(\beta(s)L')+\omega D\cosh(\beta(s)L')},
 \end{align}
 \end{subequations}
 and
 \begin{subequations}
 \begin{align}
 \widetilde{P}(x,\ell,s|0)&=\frac{\cosh(\beta(s)(L-x))}{D\cosh(\beta(s)L)}\exp\left [-\ell\beta(s)\tanh(\beta(s)L)\right],\\ \widetilde{Q}(x,\ell,s|0)&=\frac{\cosh(\beta(s)(L'+x))}{D\cosh(\beta(s)L')}\exp\left [-\ell\beta(s)\tanh(\beta(s)L')\right].
 \end{align}
\end{subequations}
The corresponding marginal probability densities are thus
 \begin{subequations}
 \begin{align}
 \widetilde{p}_+^{\Psi}(x,s|0^+)&=\frac{\cosh(\beta(s)(L-x))}{D\cosh(\beta(s)L)}\widetilde{\Psi}^+\left [\beta(s)\tanh(\beta(s)L)\right],\\ \widetilde{p}_-^{\Psi}(x,s|0^-)&=\frac{\cosh(\beta(s)(L'+x))}{D\cosh(\beta(s)L')}\widetilde{\Psi}^-\left [\beta(s)\tanh(\beta(s)L')\right].
 \end{align}
\end{subequations}
Similarly, the flux densities $\widetilde{j}_{\pm}^{\Psi}(x,s|0)$ are obtained by replacing $\widetilde{\Psi}^{\pm}$ with $D\widetilde{\psi}^{\pm}$, where
\begin{equation}
\widetilde{\psi}^{\pm}(s)=1-s\widetilde{\Psi}^{\pm}(s).
\end{equation}

The 1D version of equation (\ref{rhojay}) for $p^{\Psi}(x,t)$ on $x\in {\mathbb G}$ is
\begin{align}
p^{\Psi}(x,t)&=\Theta(x)\int_{0}^{L}\left [\int_0^{\infty} \Psi^+(\ell) {P}(x,\ell,t|x_0)d\ell\right ]h(x_0)dx_0\nonumber \\
&\quad +\Theta(-x)\int_{-\infty}^{0}\left [\int_0^{\infty}\Psi^-(\ell){Q}(x,\ell,t|y_0)d\ell\right ]\overline{g}(y_0)dy_0.
\end{align}
Similarly, the 1D version of the first renewal equation (\ref{Frenewal}) takes the form
 \begin{align}
  \label{1DFrenewal}
 \rho^{\Psi}(x,t)&=p^{\Psi}(x,t)+\frac{1}{2}\int_0^t [\rho^{\Psi}(x,t-\tau |0^+) +\rho^{\Psi}(x,t-\tau|0^- )]f^{\Psi}(\tau)d\tau 
   \end{align}
   for $x\in {\mathbb G}$.
Laplace transforming with respect to time $t$ then gives
 \begin{align}
  \label{1DFrenewal2}
 \widetilde{\rho}^{\Psi}(x,s) = \p^{\Psi}(x,s)+\frac{1}{2}[\widetilde{\rho}^{\Psi}(x,s |0^+)+\widetilde{\rho}^{\Psi}(x,s |0^-) ]\widetilde{f}^{\Psi}(s),\ x\in {\mathbb G}.
 \end{align}
 The factor $\widetilde{\rho}^{\Psi}(x,s |0^+)+\widetilde{\rho}^{\Psi}(x,s |0^-)$ can now be determined by setting $g(x_0)=[\delta(x_0-0^+) +\delta(x-0^-)]/2$ in equation (\ref{1DFrenewal2}):
   \begin{align*}
\widetilde{\rho}^{\Psi}(x,s |0^+)+\widetilde{\rho}^{\Psi}(x,s |0^-)&=\Theta(x)\p_+^{\Psi}(x,s|0^+)+\Theta(-x)\p_-^{\Psi}(x,s|0^-)\\
&+\frac{1}{2}[\widetilde{\rho}^{\Psi}(x,s |0^+)+\widetilde{\rho}^{\Psi}(x,s |0^-) ]\bigg [ \widetilde{j}_+^{\Psi}(0^+,s|0^+)+\widetilde{j}_-^{\Psi}(0^-,s|0^-)\bigg ] 
 \end{align*}
 which can be arranged to yield the result
    \begin{align*}
\widetilde{\rho}^{\Psi}(x,s |0^+)+\widetilde{\rho}^{\Psi}(x,s |0^-)= \frac{\Theta(x)\p_+^{\Psi}(x,s|0^+)+\Theta(-x)\p_-^{\Psi}(x,s|0^-)}{1-[\widetilde{j}_+^{\Psi}(0^+,s|0^+)+\widetilde{j}_-^{\Psi}(0^-,s|0^-)]/2}.
 \end{align*}
  Substituting back into equations (\ref{Frenewal2}) yields the explicit solution
 \begin{align}
  \label{Frenewal3}
 \widetilde{\rho}^{\Psi}(x,s) = \p^{\Psi}(x,s)+ \frac{\Theta(x)\p_+^{\Psi}(x,s|0^+)+\Theta(-x)\p_-^{\Psi}(x,s|0^-)}{1-[\widetilde{j}_+^{\Psi}(0^+,s|0^+)+\widetilde{j}_-^{\Psi}(0^-,s|0^-)]/2}\widetilde{f}^{\Psi}(s),\ x\in {\mathbb G}.
 \end{align}
 
 Further simplification occurs if we take $\Psi^+=\Psi^-$ and $L,L'\rightarrow \infty$, such that
 \begin{align}
 \p_{\pm}^{\Psi}(x,s|0)=\frac{1}{D}\e^{-\beta(s)|x|}\widetilde{\Psi}(\beta(s)),\quad \widetilde{j}_{\pm}^{\Psi}(x,s|0)=\e^{-\beta(s)|x|}\widetilde{\psi}(\beta(s)),
 \end{align}
 We then find that
 \begin{align}
  \label{Frenewal4}
 \widetilde{\rho}^{\Psi}(x,s) = \p^{\Psi}(x,s)+ \frac{\e^{-\beta(s)|x|}}{ 2\sqrt{sD}}\Gamma^{\Psi}(s),\ x\in {\mathbb G},
 \end{align}
 where
 \begin{align}
 \Gamma^{\Psi}(s)&\equiv \widetilde{f}^{\Psi}(s) =D\partial_x\p^{\Psi}(0^+,s)- D\partial_x\p^{\Psi}(0^-,s)\nonumber \\
 &=\widetilde{\psi}(\beta(s))  \left [\int_{0}^{\infty}\e^{-\beta(s)x_0}h(x_0)dx_0+\int_{-\infty}^{0}\e^{\beta(s)x_0}\overline{h}(x_0)dx_0\right ].
 \label{cool}
\end{align}

Since the propagator satisfies the diffusion equation in the bulk of the domain, the density $\rho^{\Psi}(x,t)$ does too. The remaining issue concerns the boundary condition at the interface. We proceed along the lines of Ref. \cite{Bressloff22}. First, it follows from equation (\ref{Frenewal4}) that
\begin{subequations}
 \label{enc0}
 \begin{align}
  \widetilde{\rho}^{\Psi}(x,s) + \widetilde{\rho}^{\Psi}(-x,s)&= \p^{\Psi}(x,s)+\p^{\Psi}(-x,s)+ \frac{ \e^{-\beta(s)|x|}}{\sqrt{s D}}\Gamma^{\Psi}(s),\\
   \widetilde{\rho}^{\Psi}(x,s) - \widetilde{\rho}^{\Psi}(-x,s)&= \p^{\Psi}(x,s)-\p^{\Psi}(-x,s),
 \end{align}
 \end{subequations}
 and
\begin{subequations}
\label{enc1}
 \begin{align}
&D \partial_x\widetilde{\rho}^{\Psi}(0^+,s) - D\partial_x\widetilde{\rho}^{\Psi}(0^-,s) = D\partial_x\p^{\Psi}(0^+,s)- D\partial_x\p^{\Psi}(0^-,s)- \Gamma^{\Psi}(s) ,\\
& D\partial_x\widetilde{\rho}^{\Psi}(0^+,s) +D \partial_x\widetilde{\rho}^{\Psi}(0^-,s) = D\partial_x\p^{\Psi}(0^+,s)+ D\partial_x\p^{\Psi}(0^-,s).
 \end{align}
 \end{subequations}
Equations (\ref{cool}) and (\ref{enc1}a) establish that $ \partial_x\widetilde{\rho}^{\Psi}(0^+,s) = \partial_x\widetilde{\rho}^{\Psi}(0^-,s)$. In other words, the flux through the membrane is continuous, as it is in the standard permeable boundary conditions (\ref{1Dmaster}c,d). Equation (\ref{enc1}b) then implies that
\begin{align}
2D \partial_x\widetilde{\rho}^{\Psi}(0^{\pm},s) &=\widetilde{\psi}(\beta(s))\left [\int_{0}^{\infty}\e^{-\beta(s)x_0}h(x_0)dx_0-\int_{-\infty}^0\e^{\beta(s)x_0}\overline{h}(x_0)dx_0\right ]\nonumber \\
&=\frac{D \widetilde{\psi}(\beta(s))}{\widetilde{\Psi}(\beta(s))}[\p^{\Psi}(0^+,s)-\p^{\Psi}(0^-,s)]\nonumber \\
&=\frac{D \widetilde{\psi}(\beta(s))}{\widetilde{\Psi}(\beta(s))}[\widetilde{\rho}^{\Psi}(0^+,s)-\widetilde{\rho}^{\Psi}(0^-,s)].
\end{align}
The final line follows from equation (\ref{enc0}b).

\begin{figure}[b!]
  \centering
  \includegraphics[width=8cm]{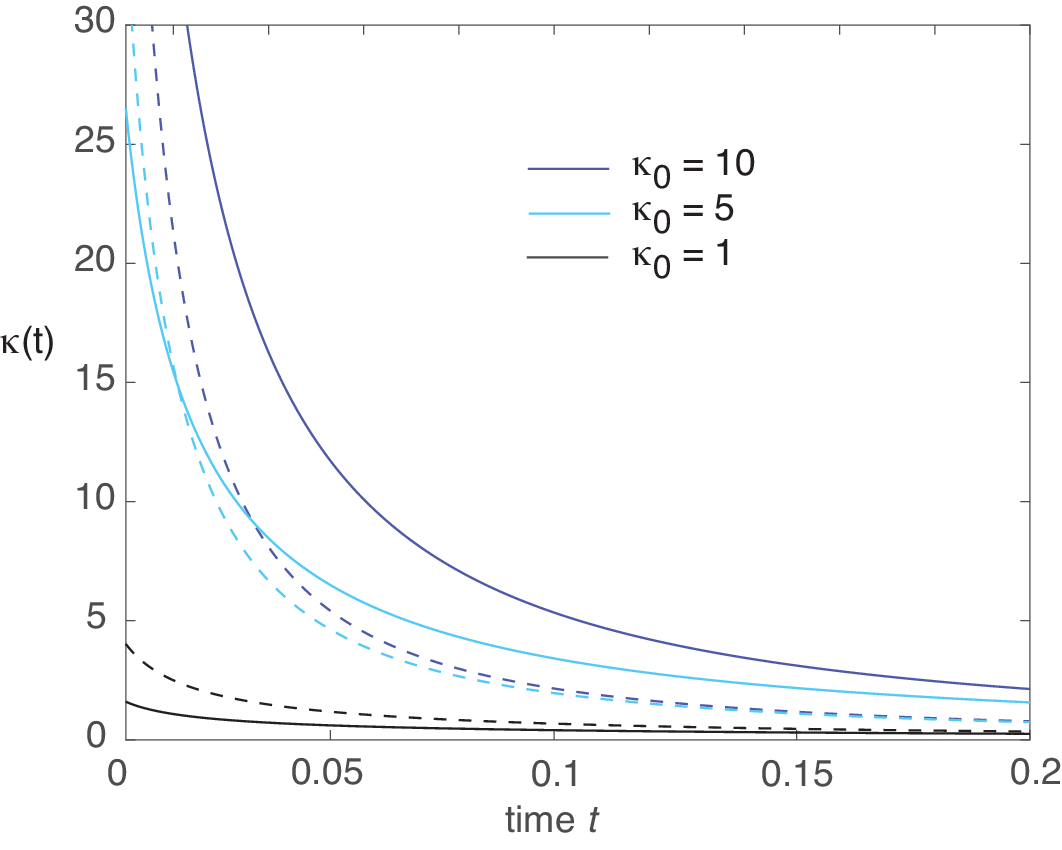}
  \caption{Plot of permeability function $\kappa(t)$ as a function of time $t$ for various values of $\omega_0$ with $D=10$ (solid curves) and $D=1$ dashed curves.}
  \label{fig5}
\end{figure}
 
 In the exponential case $\Psi(\ell)=\e^{-\omega_0\ell}$, we have $\psi(\ell)=\omega_0 \Psi(\ell)$, and we recover the semipermeable boundary conditions of equations (\ref{1Dmaster}c,d) with permeability $\kappa=\omega_0D$ and $\alpha=1/2$. For non-exponential distributions, the boundary condition involves a time-dependent permeability. More specifically, setting
 \begin{equation}
 \label{Lkap}
 \widetilde{\kappa}(s)=\frac{D \widetilde{\psi}(\beta(s))}{\widetilde{\Psi}(\beta(s))}
 \end{equation}
 and using the convolution theorem, the boundary condition in the time domain takes the form
 \begin{align}
2D \partial_x {\rho}^{\Psi}(0^{\pm},t) 
&=\int_0^t \kappa(\tau)[ {\rho}^{\Psi}(0^+,t-\tau)- {\rho}^{\Psi}(0^-,t-\tau)]d\tau.
\end{align}
For the sake of illustration, suppose that $\psi(\ell) $ is given by the gamma distribution:
\begin{equation}
\label{psigam}
\psi(\ell)=\frac{\omega_0(\omega_0 \ell)^{\mu-1}\e^{-\omega_0 \ell}}{\Gamma(\mu)}, \quad \mu >0,
\end{equation}
where $\Gamma(\mu)$ is the gamma function. Here $\omega_0$ determines the effective absorption rate and $\mu $ characterizes the deviation of $\psi(\ell)$ from the exponential case $\mu=1$. The corresponding Laplace transforms are
\begin{equation}
\widetilde{\psi} (z)=\left (\frac{\omega_0}{\omega_0+z}\right )^{\mu},\quad \widetilde{\Psi}(z)=\frac{1-\widetilde{\psi}(z)}{z}
\end{equation}
 If $\mu <1$ ($\mu>1$) then $\psi(\ell)$ decreases more rapidly (slowly) as a function of the local time $\ell$. Substituting the gamma distribution into equation (\ref{Lkap}) yields
 \begin{equation}
 \widetilde{\kappa}(s)=\frac{\sqrt{sD}\omega_0^{\mu}}{(\omega_0+\beta(s))^{\mu}-\omega_0^{\mu}}.
 \end{equation}
 If $\mu=1$ then $\widetilde{\kappa}(s)=\omega_0 D=\kappa_0$ and $\kappa(\tau)=\omega_0\delta(\tau)$. An example of $\mu\neq 1$ that has a simple inverse Laplace transform is $\mu=2$:
  \begin{equation}
 \widetilde{\kappa}(s)=\frac{D\omega_0^2}{2 \omega_0+\beta(s)}=\frac{\kappa_0^2 }{2\kappa_0 +\sqrt{sD}}
 \end{equation}
 and
 \begin{align}
 \kappa(\tau)=\frac{\kappa_0^2}{\sqrt{D}}\left [\frac{1}{\sqrt{\pi \tau}}-\frac{2\kappa_0}{\sqrt{D}}\e^{4\kappa_0^2\tau/D} \mbox{erfc}(2\kappa_0\sqrt{\tau/D})\right ],
 \label{kaptp}
 \end{align}
 where $\mbox{erfc}(x)=(2/\sqrt{\pi})\int_x^{\infty} \e^{-y^2}dy$ is the complementary error function. Example plots of $\kappa(\tau)$ are shown in Fig. \ref{fig5}. It can be seen that $\kappa$ is a monotonically decreasing function of time whose rate of decay depends on $\kappa_0$ and $D$.
Asymptotically expanding $\mbox{erfc}(x)$ in equation (\ref{kaptp}) using the formula
 \begin{equation}
 \mbox{erfc}(x)\sim \frac{1}{\sqrt{\pi}}\e^{-x^2} \sum_{k=0}^{\infty} (-1)^k\frac{(2k)!}{2^{2k}k!}\frac{1}{x^{2k+1}},
 \end{equation}
 shows that $\kappa(t)$ is heavy-tailed with
  \begin{equation}
 \kappa(t)\sim \sqrt{\frac{D}{\pi}} \frac{1}{8t^{3/2}} ,\ t \rightarrow \infty,
 \end{equation}

   \subsection{Absorbing target}
   
 So far in our discussion of snapping out BM we have ignored absorption within the target $\calU$. If absorption is included, then each round of partially reflected BM within $\calU$ has two distinct killing events: either the particle is first absorbed on $\partial \calU$, after which snapping out BM continues in the normal fashion, or the particle is first absorbed within the interior $\calU$, after which the stochastic process is terminated. A single round of BM within $\calU$ thus becomes a competition between two absorbing targets, namely, the boundary $\partial \calU$ and the interior $\calU$. Certain care has to be taken in treating these two targets as independent, since $\calU$ is an open set whose closure includes $\partial \calU$. One way to deal with this situation would be to introduce a boundary layer around $\partial \calU$, within which the particle can only be absorbed by $\partial \calU$. This is also consistent with how one would numerically calculate the boundary local time. Here we ignore such details, and simply consider the dual-aspect propagator $Q(\x,\ell,a,t|\x_0)$ for the pair $(\ell_t^-,\calA_t)$ with $\calA_t$ the occupation time within $\calU$, see equation (\ref{occ}).

The evolution equation for $Q(\x,\ell,a,t|\x_0)$ is obtained by combining  equations (\ref{Qloc}a) and (\ref{Pocc}c):
\begin{subequations} 
\label{Qocloc}
\begin{align}
 &\frac{\partial Q(\x,\ell,a,t|\x_0)}{\partial t}+\frac{\partial Q(\x,\ell,a,t|\x_0)}{\partial a}=D\nabla^2 Q(\x,\ell,a,t|\x_0)  -\delta(a)Q(\x,\ell,a=0,t|\x_0), \nonumber \\
&\hspace{6cm} \x \in   \calU, \\
 &D\nabla Q(\x,\ell,a,t|\x_0) \cdot \n_0= D Q(\x,\ell=0,a,t|\x_0) \ \delta(\ell)  +D\frac{\partial}{\partial \ell} Q(\x,\ell,a,t|\x_0),  \x\in \partial \calU.
\end{align}
\end{subequations}
Let $\Psi_1(\ell)$ and $\Psi_2(a)$ denote the threshold distributions for the local time and occupation time, respectively. Following along similar lines to our previous examples, the associated marginal probability density is
\begin{align}
q^{\Psi_1,\Psi_2}(\x,t|\x_0)&=\int_0^{\infty}d\ell\, \Psi_{1}(\ell) \int_0^{\infty} da\, \Psi_2(a) Q(\x,\ell,a,t|\x_0)\nonumber \\
  \label{oo}
 &=\int_0^{\infty}d\ell\, \Psi_{1}(\ell) \int_0^{\infty} da\, \Psi_2(a){\rm LT}_{\omega}^{-1}{\rm LT}_{z}^{-1}\widetilde{Q}(\x,\omega,z,t|\x_0),
  \end{align}
  with
  \begin{subequations}
\label{QoclocLT}
\begin{align}
&\frac{\partial \widetilde{Q}(\x,\omega,z,t|\x_0)}{\partial t}=D\nabla^2 \widetilde{Q}(\x,\omega,z,t|\x_0) -\gamma \widetilde{Q}(x,\omega,z,t|x_0),\\
&\nabla\widetilde{Q}(\x,\omega,z,t|\x_0)\cdot \n_0=\omega \widetilde{Q}(\x,\omega,z,t|\x_0),\quad \x \in \partial \calU.
\end{align}
\end{subequations}
The corresponding marginal fluxes are as follows:
  \begin{align}
 {J}_1^{\Psi_1,\Psi_2}(\x_0,t) &\equiv D \int_{\partial \calU} \nabla q^{\Psi_1,\Psi_2}(\x,t|\x_0)\cdot \n_0d\x  \nonumber \\
 &=D \int_{\partial \calU}\left \{ \int_0^{\infty}d\ell\, \psi_1(\ell) \int_0^{\infty} da\, \Psi_2(a) Q(\x,\ell,a,t|\x_0)\right \}d\x,
  \label{fm}\\
  \label{fp}
  {J}_2^{\Psi_1,\Psi_2}(\x_0,t) &=D \int_{ \calU}\left \{ \int_0^{\infty}d\ell\, \Psi_1(\ell) \int_0^{\infty} da\, \psi_2(a) Q(\x,\ell,a,t|\x_0)\right \}d\x.
  \end{align}

The stochastic process is killed as soon as one of the contact times exceeds its corresponding threshold, which occurs at the stopping time $\calT=\min\{\tau_{1},\tau_2\}$ with
\begin{equation}
 \tau_{1} =\inf\{t>0: \ell_t>\widehat{\ell} \}, \quad \tau_2=\inf\{t>0: \calA_t>\widehat{\calA} \}.
\label{goat}
\end{equation}
Since there are now two effective targets, we have to introduce the associated splitting probabilities $\pi_j^{\Psi_1,\Psi_2}(\x_0)$ and conditional FPT densities $f_j^{\Psi_1,\Psi_2}(\x_0,t)$ for $j=1,2$. These are defined according to
\begin{align}
\pi_j^{\Psi_1,\Psi_2}(\x_0)&=\int_0^{\infty}{J}_j^{\Psi_1,\Psi_2}(\x_0,t) dt,\quad
f_j^{\Psi_1,\Psi_2}(\x_0,t)=\frac{{J}_j^{\Psi_1,\Psi_2}(\x_0,t)}{\pi_j^{\Psi_1,\Psi_2}(\x_0) }.
\end{align}
We can now define snapping out BM with absorption using the conditional FPTs. In particular, the first renewal equation
still holds with $\Psi\rightarrow (\Psi_1,\Psi_2)$ and the FPT density $f^{\psi}$ replaced by $\pi_j^{\Psi_1,\Psi_2} f_j^{\Psi_1,\Psi_2}$.

\setcounter{equation}{0}

\section{Discussion} In this paper we considered the FPT problem for a single target $\calU$ in a bounded domain whose interior is partially absorbing and whose boundary $\partial \calU$ is a semi-permeable interface, see Fig. \ref{fig1}(b). We described several scenarios of increasing complexity.
\begin{enumerate}

\item A classical semi-permeable membrane $\calU$ with permeability $\kappa$ and bias $\alpha$, and a constant rate of absorption $\gamma$ within $\calU$. We showed that one way to solve the FPT problem was in terms of the spectral properties of a pair of D-to-N operators.
\medskip

\item A classical semi-permeable membrane $\calU$ and a non-Markovian process of absorption within $\calU$. We used an encounter-based method to formulate the absorption process in terms of a random threshold-crossing condition for the occupation time within $\calU$. If the probability density of the occupation time threshold is an exponential, then one recovers the case of a constant rate of absorption. On the other hand, a non-exponential density leads to a non-Markovian form of absorption. The resulting MFPT depends on various moments of the occupation time threshold.
\medskip

\item It is also possible to generalize the classical model of a semi-permeable membrane by formulating single-particle diffusion in terms of snapping out BM. Snapping out BM latter sews together successive rounds of partially reflecting BM that are restricted to either the interior or the exterior of $\partial \calU$. Each round of partially reflected BM is killed when the boundary local time on the current side of the semi-permeable interface exceeds a randomly generated local time threshold. If the probability density of the latter is exponential, then the classical case of constant permeability is recovered. On the other hand, a non-exponential density leads to a time-dependent permeability that tends to be heavy-tailed. It is also possible to combine snapping out BM with a non-Markovian absorption mechanism by keeping track of both the boundary local time on $\partial \calU^-$ and the occupation time within the interior $\calU$.
\end{enumerate}

There are a number of natural generalizations of the single target problem considered here.
\begin{enumerate}

\item[(i)] {\em Multiple targets with semipermeable interfaces in $\R^d$}. As we highlighted in this paper, solving the FPT problem for a single target in 2D or 3D is non-trivial even for simplified geometries. The analysis becomes even more difficult in the case of multiple targets, where one has to calculate splitting probabilities and conditional FPTs. However, considerable simplification occurs in the small-target limit, since one can then use matched asymptotic expansions and Green's function methods. More specifically, an inner or local solution is constructed in an $O(\epsilon)$ neighborhood of each target, where $\epsilon$ characterizes the relative size of each target compared to the size of the search domain. (The inner solution ignores the effects of other targets and treats the search domain as $\R^d$.) The inner solution is then matched to an outer or global solution that is valid away from each neighborhood. For details see Ref. \cite{Bressloff23a} for 2D and  Ref. \cite{Bressloff23b} for 3D.

\medskip

\item[(ii)] {\em Numerical methods.} In this paper, we focused on analytical methods for solving the target problem with semi-permeable interfaces. If one is also interested in studying single-particle trajectories, then it is necessary to construct efficient numerical schemes for simulating snapping out BM. Along these lines, we have recently developed a fast Monte Carlo algorithm for solving multi-dimensional snapping out BM for multiple interfaces, which combines a walk-on-spheres method with an efficient numerical scheme for calculating boundary local times \cite{Schumm23}. The numerical methods were shown to have high accuracy when compared to solutions obtained from matched asymptotic analysis in the small-target regime. Note that there are also a number of alternative computational schemes for solving 1D diffusion problems in heterogeneous media with semi-permeable interfaces \cite{Farago18,Farago20,grebenkov-sp}. However, these do not generate exact sample trajectories of snapping out BM.

\medskip

\item[(iii)] {\em Biophysical mechanisms.} Finally, from a modeling perspective, it would be interesting to identify plausible biophysical mechanisms underlying non-Markovian models of semi-permeable membranes. It is known that various surface-based reactions are better modeled in terms of a reactivity that is a function of the local time. 
For example, the surface may become progressively
activated by repeated encounters with a diffusing
particle, or an initially highly reactive surface may become less active due to multiple interactions with the particle (passivation) \cite{Bartholomew01,Filoche08}. One potential application is synaptic receptor trafficking in neurons \cite{Bressloff23a}, where the clustering of receptors within postsynaptic domains can be modeled in terms of a diffusion-trapping model. In this example, the boundary of the postsynaptic domain could be treated as an asymmetric semipermeable membrane that is likely to involve non-Markovian components due to the complexity of the crowded molecular environment.

\end{enumerate}

\end{document}